\documentclass[onecolumn]{emulateapj}

\shorttitle{The 2015 Summer Solstice event}
\shortauthors{Augusto et al.}

\begin{document}


\title
{The 2015 Summer Solstice Storm: one of the major geomagnetic storms of solar cycle 24 observed at ground level}


\author{C. R. A. Augusto,  C. E. Navia and M. N. de Oliveira\altaffilmark{1}}
\affil{Instituto de F\'{i}sica, Universidade Federal Fluminense, 24210-346, Niter\'{o}i, RJ, Brazil}

\author{A. A. Nepomuceno}
\affil{Departamento de Ci\^{e}ncias da Natureza, Universidade Federal Fluminense, 28890-000, Rio das Ostras, RJ, Brazil}

\author{J. P. Raulin, E. Tueros}
\affil{Centro de R\'{a}dio Astronomia e Astrof\'{i}sica Mackenzie (CRAAM), Universidade Presbiteriana Mackenzie, 01302-907, S\~{a}o Paulo, SP, Brazil}

\author{R. R. S. de Mendon\c{c}a}
\affil{State Key Laboratory of Space Weather, National Space Science Center (NSSC), Chinese Academy of Sciences, NO. 1 Nanertiao, Zhongguancun, Beijing, 100190, China\\
Space Geophysics Division, National Institute for Space Research (INPE), 12227-010, S\~{a}o Jos\'{e} dos Campos, SP, Brazil}

\author{A. C. Fauth, H. Vieira de Souza}
\affil{Instituto de F\'{i}sica Gleb Wathagin, Universidade Estadual de Campinas, 13083-859, Campinas, SP, Brazil}

\author{V. Kopenkin}
\affil{Research Institute for Science and Engineering, Waseda University, Shinjuku, Tokyo 169, Japan}

\author{T. Sinzi}
\affil{Rikkyo University, Toshima-ku, Tokyo 171, Japan}


\altaffiltext{1}{E-mail address:paulista@fisica.if.uff.br}


\begin{abstract}
We report on the 22-23 June 2015 geomagnetic storm which occurred during the summer solstice time. There has been a shortage of intense geomagnetic storms during the current solar cycle 24 in relation to the previous cycle. This situation changed after mid-June 2015 when one of the biggest solar active regions (AR 2371) of current solar cycle 24, close to the central meridian produced several coronal mass ejections (CMEs) associated with M-class flares. The CMEs impact on the Earth's magnetosphere resulted in a moderately-severe G4-class geomagnetic storm on 22-23 June 2015 and a G2 (moderate) geomagnetic storms on 24 June. The G4 solstice storm was the second biggest (so far) geomagnetic storms of cycle 24. We highlight the ground level observations made by New-Tupi, Muonca and the CARPET El Leoncito cosmic ray detectors that are located within the South Atlantic Anomaly (SAA) region. These observations are studied in correlation with data obtained by space-borne detectors (ACE, GOES, SDO, and SOHO) and other ground-based experiments. The CME designations are from the Computer Aided CME Tracking (CACTus) automated catalog. As expected, Forbush Decreases (FD) associated with the passing CMEs were recorded by these detectors. We noticed a peculiar feature linked to a severe geomagnetic storm event. The 21 June 2015 CME 0091 was likely associated with the 22 June summer solstice FD event. The angular width of CME 0091 was very narrow and measured $\sim 56^{\circ}$ degrees seen from Earth. In most cases, only CME halos and partial halos, lead to severe geomagnetic storms. We performed a cross-check analysis of the FD events detected during the rise phase of the current solar cycle 24, the geomagnetic parameters, and the CACTus CME catalog. Our study suggests that narrow angular-width CMEs that erupt in the western region of the ecliptic plane can lead to moderate and severe geomagnetic storms. We also report on the strong solar proton radiation storm with onset on 21 June. We did not find a signal from this SEP at ground level. The details of these observations are presented.
\end{abstract}



\keywords{sun:activity, astroparticle physics, atmospheric effects, instrumentation:detectors}


\section{Introduction}
\label{sec:intro}

\begin{figure}
\vspace*{-0.0cm}
\hspace*{-0.0cm}
\centering
\includegraphics[width=12.0cm]{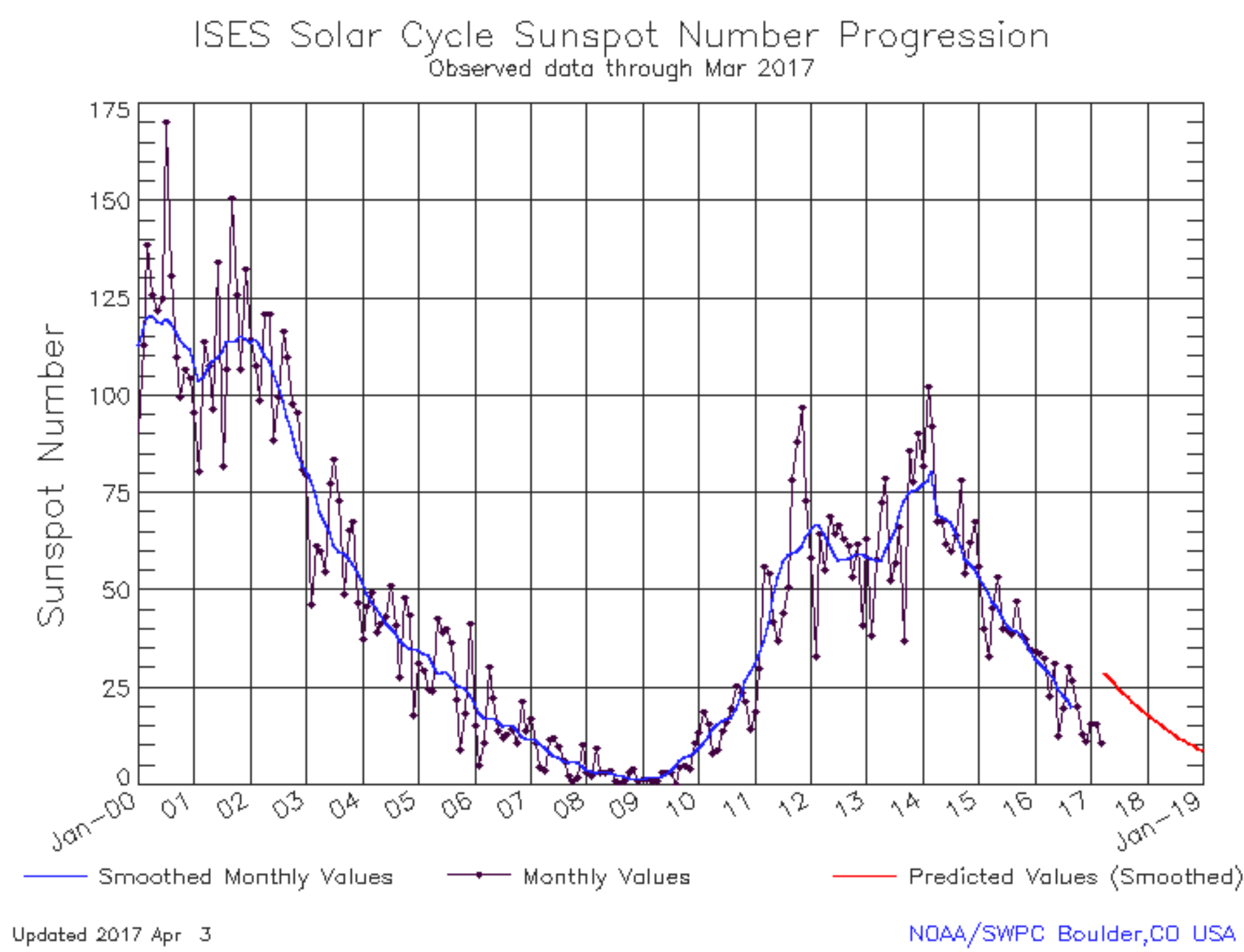}
\vspace*{-0.0cm}
\caption{Solar Cycle Sunspot Number Progression. Updated 2017 April 3. The graph reports the maximum period from the solar cycle 23 to the current situation of the descending phase of the solar cycle 24 (with data until March 2017). We can note that the solar cycle 24 had two peaks, with the second peak being larger than the first. The black line represents the monthly averaged data and the blue line represents a 13-month smoothed version of the monthly averaged data. The official International Sunspot Number (RI) is issued by the Sunspot Index Data Center (SIDC) in Brussels, Belgium. This plot displays the SIDC monthly sunspot numbers. Credit: NOAA/SWPC.}
\vspace*{-0.0cm}
\label{fig1}
\end{figure}

Since its inception in 2008, the current solar cycle 24 has been weak compared with prior periods in recent cycles. Historically, sunspots (intense magnetic activity areas that appear as dark spots compared to surrounding regions) are indicators of the nearly periodic 11-year solar magnetic activity cycle\footnote{Cycle 1 starts in 1755 after Rudolph Wolf created a standard sunspot number index. Solar maximum and solar minimum refer respectively to periods of maximum and minimum sunspot counts.}. The current cycle 24 has two peak maximum of sunspots as shown in Fig.~\ref{fig1}. In the current cycle 24 the second peak\footnote{\url{http://solarscience.msfc.nasa.gov/predict.shtml}} (in 2014) in sunspot number was larger than the first (in 2011).

The cycle 24 has the smallest maximum monthly sunspot number since cycle 14, which took place between February 1902 and August 1913. It could be an indication of the beginning of some change in solar variability \citep{pes16}. While the most likely outcome of the present changes in the open magnetic flux of the Sun is a fall to average conditions, statistical analysis showed that there is an $8\%$ chance of a return to Maunder Minimum\footnote{During the Maunder Minimum (1645-1715) sunspots became exceedingly rare.}within the next 40 years \citep{loc10,ine15}. Earth's interplanetary environment is constantly influenced by the transient disturbances propagating from the Sun. Interplanetary counterparts of coronal mass ejections (CMEs) are often associated with solar flares and active regions on the Sun's surface \citep{bur95,gon90,tbt92,wdg99}.

If the interplanetary counterparts of the CMEs have a significant southward component (Bz), of the interplanetary magnetic field (IMF), in either the sheath behind the shock or in the driver gas (magnetic cloud), then after reaching Earth's magnetosphere, they may lead to geomagnetic storms \citep{tbt88,tbt92}.

\begin{figure}
\vspace*{-0.0cm}
\hspace*{-0.0cm}
\centering
\includegraphics[width=8.0cm]{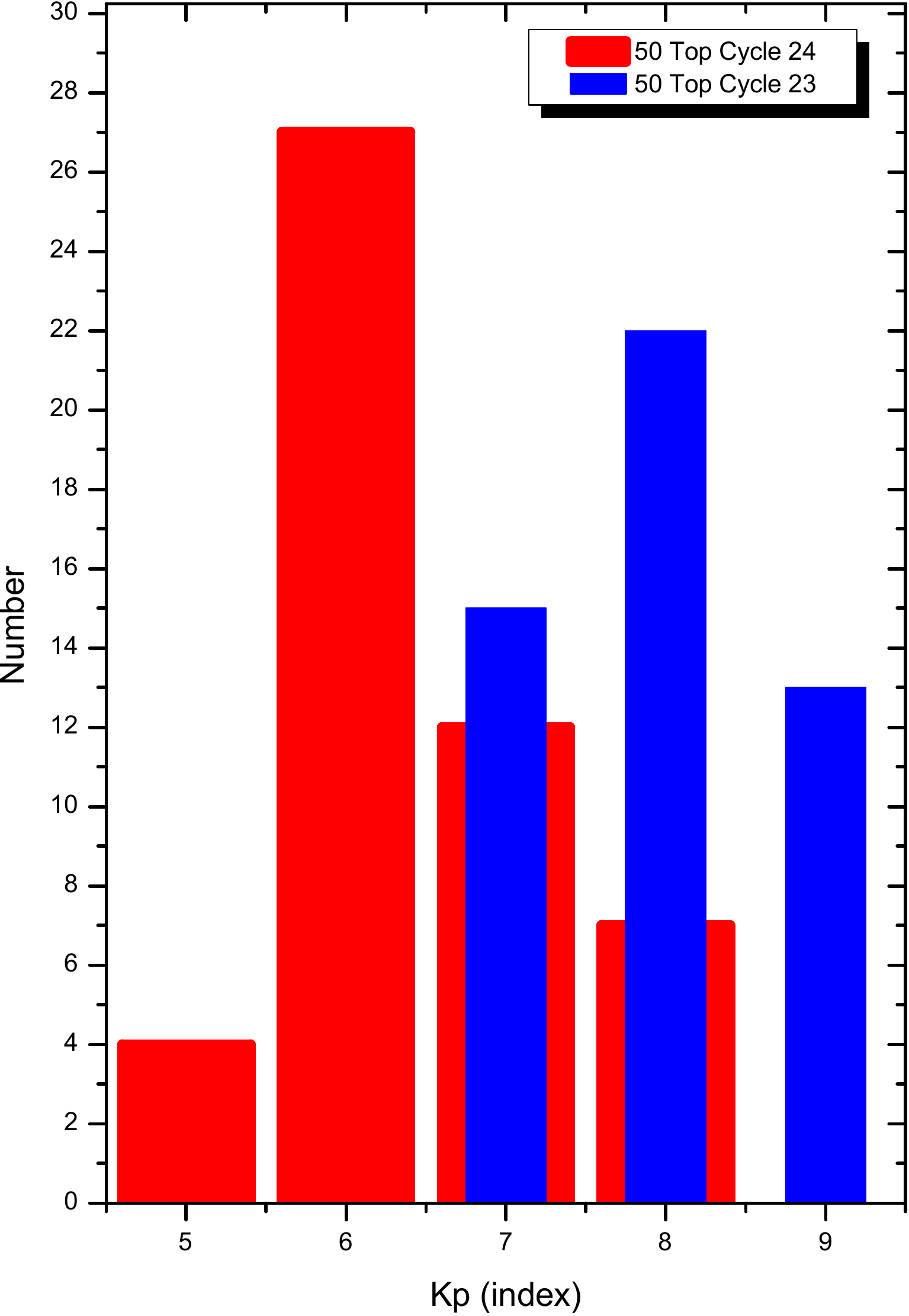}
\vspace*{-0.0cm}
\caption{The planetary Kp index distribution of the top 50 strongest geomagnetic storms in cycle 24 (the red histogram) and cycle 23 (the blue histogram).
}
\label{fig2}
\end{figure}

The frequency of geomagnetic storms varies with the solar cycle \citep{gon90}. The mild space weather during cycle 24 can be illustrated by the number of major geomagnetic storms in comparison with its predecessor, cycle 23. In Fig.~\ref{fig2} we show the Kp index\footnote{The estimated 3 hr planetary Kp index is derived from the National Oceanic and Atmospheric Administration (NOAA) Space Weather Prediction Center (SWPC) using data from the ground-based magnetometers.} distribution of the top 50 strongest geomagnetic storms in cycle 23\footnote{\url{https://www.spaceweatherlive.com/en/auroral-activity/top-50-geomagnetic-storms/solar-cycle/23}} \citep{Zhang07} and cycle 24\footnote{\url{https://www.spaceweatherlive.com/en/auroral-activity/top-50-geomagnetic-storms/solar-cycle/24}} (as of 2016 September) \citep{Pande18}. It is clear that the number of major storms has declined in cycle 24 relative to cycle 23. There were observed 22 events with $Kp=8$ (G4, severe)\footnote{\url{http://www.swpc.noaa.gov/noaa-scales-explanation}} in cycle 23, and 7 events of this category in cycle 24. There is a decline in the number of the strongest geomagnetic storms, those cataloged as $Kp=9$ (G5, extreme): 13 events in cycle 23 and none (so far) in cycle 24. The majority of the strongest geomagnetic storms in cycle 24 are of $Kp=6$ (G2, moderate). Many peculiarities and anomalies of cycle 24 and their effects on Earth have been the subject of multiple studies \citep[e.g.][]{gop14,gop15}.

Solar disturbances affect the propagation of galactic cosmic rays. When the solar disturbances pass the Earth's magnetosphere, the ground-based detectors can observe depressions of cosmic ray intensity, the so-called Forbush Decreases (FD) \citep{for37}. Typically, the bow shock at the front of the magnetic cloud (ICME) can act as a barrier that prevents cosmic rays from reaching the Earth. The decrease in the observed intensity is followed by a gradual recovery that typically lasts up to several days \citep{loc71}. Most of the cosmic-ray decreases observed by ground-based neutron monitors are attributed to CME-driven geomagnetic storms \citep{can96,can00,can03}. The magnitude of an FD observed by a particular detector depends on several factors, such as the size of the CME, the strength of the magnetic fields in the CME, the proximity of the CME to the Earth, and the location of the detector on the Earth \citep{cli96,kim08}.

In this work, we present the first results of the analysis of data acquired by cosmic ray detectors located in the South Atlantic Anomaly (SAA) region during one of the strongest (so far) geomagnetic storms of cycle 24, the 22-23 June 2015 geomagnetic storm. Different aspects of the storm, that occurred during the 2015 summer solstice\footnote{The June solstice is the shortest day in the southern hemisphere and the longest day in the north.} time, were previously reported by other groups \citep{liu15,ast16,moh16,bak16}.

We present and analyze the FD associated with this geomagnetic storm using data from New Tupi and Muonca muon telescopes located in Brazil and the CARPET (El Leoncito) detector located in Argentina. This event is also studied in correlation with observations reported by spacecraft detectors (ACE, GOES, SDO, and SOHO) and other ground-based experiments.

The organization of the paper is as follows. In section~\ref{sec:experimental}, we give a brief description of the experiments and methodology. Section~\ref{sec:solstice} describes the temporal sequence of events during an active storm period of June 2015 and the Solstice Storm event. Section~\ref{sec:check} presents a cross-check between the CACTus CME catalog,  the geomagnetic parameters and the FD events detected during the ascension phase of the current solar cycle 24 (2010-2013). The aim is to characterize the CMEs triggering geomagnetic storms, associated with FD observed in ground level detectors. We include in Section~\ref{sec:radiation} the information on solar energetic particles observed in spacecraft detectors in the period. In Section~\ref{sec:conclusion}, we present our conclusions. 

The paper includes an appendix that presents a table with the basic information on the selected FDs (for the time interval 2010-2013 and the solstice event), the CME parameters from the CACTus CME catalog and the ring current $Dst$ indices. The FD parameters are based on the Oulu NM data, the FD data listed in \cite{lin16} and New-Tupi telescope data.


\section{Experimental setup and methodology}
\label{sec:experimental}

\subsection{Cosmic rays at ground level}
\label{sec:cosmic}

Cosmic rays are highly energetic particles (mostly protons) reaching Earth from all directions in space. In general, there are two categories of cosmic rays: primary and secondary. Primary cosmic rays originate far outside Earth's atmosphere. They are mainly of galactic origin, though include contributions from transient particle flux of solar origin and an anomalous component originating at the edge of the heliosphere \citep{fic01}. Secondary cosmic rays are particles produced within Earth's atmosphere due to  collisions between primary cosmic rays in the atmosphere, producing a cascade of secondary collisions and particles known as a shower, air shower, or cascade shower. Secondary cosmic rays shower down to Earth's surface and even penetrate it. The decay of pions and kaons produced in these collisions results in a great variety of species, for instance, a neutral pion decays to two photons (gamma rays) that, by pair production phenomena, are transformed in an electron-positron pair which can produce photons by Bremsstrahlung. It is known as the electromagnetic component ($\gamma$ and $e^{\pm}$) of the air shower. Ground level detectors at mountain altitudes can detect this component.  Following the charged pion decay, the shower has a muon component. The muon is an unstable subatomic particle with a mean lifetime of 2.2 microseconds. Muons lose their energy by ionization of the material through which they pass. The time dilation effect of special relativity (from the viewpoint of the Earth) allows secondary muons of high energy to reach the detector at sea level and even penetrate deep underground. Muons produced at about 10-20 km heights above Earth surface are the most numerous energetic particles ($\sim 95\%$ at $E>1$ GeV) arriving at sea level. Muon telescopes use coincidence detection method.


\subsection{Cosmic ray detectors in the South Atlantic Anomaly (SAA) region}
\label{sec:saa}

\begin{figure}
\vspace*{-1.0cm}
\hspace*{0.5cm}
\centering
\includegraphics[width=10.0cm]{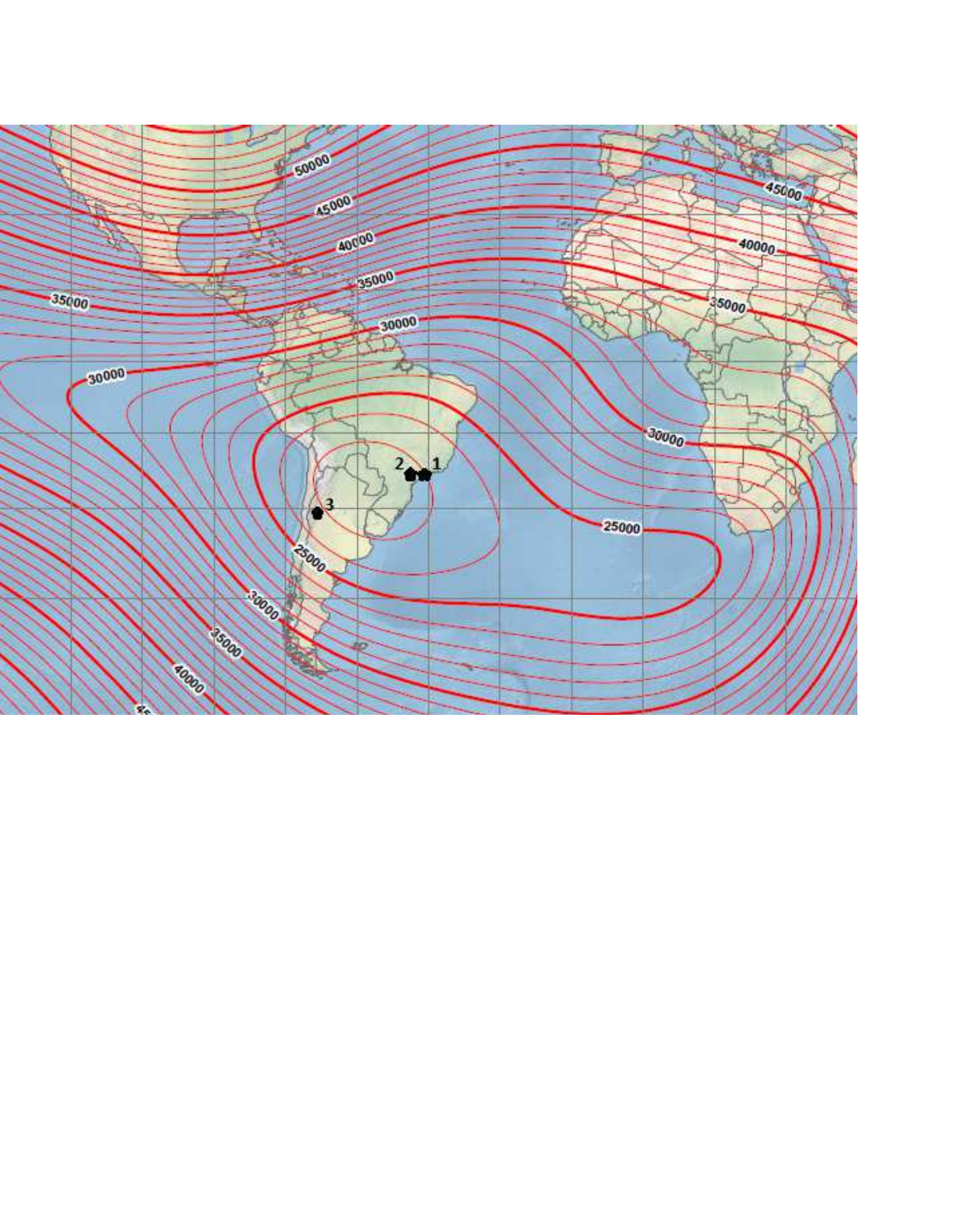}
\vspace*{-5.0cm}
\caption{Geomagnetic curves of total magnetic field intensity according to the US/UK World Magnetic Model Epoch 2000. 
The points with numbers 1, 2 and 3 indicate the physical location of New-Tupi, Muonca and El Leoncito cosmic ray detectors, respectively.}
\label{fig3}
\end{figure}

In this paper, we place an emphasis on three cosmic-ray experiments dedicated to studying space weather effects and diverse solar transient events at ground level. All of them are located in the SAA, which embraces a great part of South America's central region. The SAA is the region where the Earth's inner Van Allen radiation belt is closest to the planet's surface. The result is a deep depression in the magnetosphere. It reduces the shielding effect to charged particles that come from space \citep{caso09}. Fig.~\ref{fig3} shows geomagnetic curves of total magnetic field intensity according to the US/UK World Magnetic Model Epoch 2000 \citep{mac00} together with the location of the three cosmic rays detectors used in this work. The SAA region is clearly indicated by the lowest magnetic field over the Earth (with a field strength of $\sim 22000$ nT). However, the horizontal magnetic field components in the SAA is only slightly higher than those observed at polar regions (see Table 1). 

This behavior introduces a sub-cutoff in the geomagnetic rigidity, below the nominal Stormer cut-off, as reported by the Pamela satellite \citep{caso09}. We have shown that this effect of the SAA, seen at satellite altitudes, extends at least in part to the particles detected by ground-level instruments. Indeed, under special conditions, an excess of muons at ground level was observed in association with a high-speed stream impact \citep{aug3}.

In Fig.~\ref{fig3} the point with number 1 marks the location of  New-Tupi experiment at Niteroi city, Brazil \citep{aug3}. Number 2 shows the location of the Muonca (MUONs in Campinas) experiment designed to study the muon flux variations associated with solar events. The detector is situated in Campinas, Brazil \citep{fauth}. This experiment has been taking data continuously since 2014 April 1. The third point marks the CARPET cosmic ray detector at Complejo Astronomico El Leoncito (CASLEO)\footnote{CARPET, El Leoncito or CARPET (El Leoncito) are used interchangeably hereafter.} in Argentina that has been monitoring secondary cosmic ray intensity since 2006 \citep{leo}. Table 1 summarizes the situation and includes information on the Oulu neutron monitor used in this work.

\begin{table}[!h]
\centering
\hspace{-2.5cm}
\caption{Information about coordinates, altitude, rigidity cutoff and the components horizontal ($B_h$) and vertical ($B_v$) of the terrestrial magnetic field, on the sites of the cosmic rays experiments used in this work.}
\label{my-label}
\begin{tabular}{cccccc}
\hline
\hline
Experiment & Coordinates & Altitude (m) & R (GV) & $B_h$ (nT) 
 & $B_v$ (nT)\\
 \hline
\hline
Oulu   & $65^{\circ} 2'60''N,\; 25^{\circ} 28'12''E$                                      & 15                                                 & 0.8       & 	12716.3  & 		51411.4       \\  
\hline
New-Tupi   & $22^{\circ} 53'0''S,\; 43^{\circ}6'13''W$  & 5  & 9.2                                                                   & 	17953.6 & -14741.3       \\ 
\hline
Muonca     & $22^{\circ} 54'20''S,\; 47^{\circ} 3'39''W$      & 640                                                    & 9.2  & 18552.9& -13433.7    \\ 
\hline
El Leoncito   &  $31^{\circ} 44'11''S,\; 69^{\circ}16'39''W $                                      & 2550                                                   & 9.8       & 	19794.3  & -12544.8      \\ 

\hline
\end{tabular}
\%end{tablenotes}
\end{table}

\subsection{New-Tupi and Muonca}
\label{sec:tupi}

The muon stations located in Brazil, Muonca and New-Tupi are based respectively at the University of Campinas, S\~ao Paulo and the Fluminense Federal University, Niteroi, Rio de Janeiro. The distance between these two cities is 480 km. Each muon station has two muon telescopes, and each telescope consists of two detectors. The aim of the muon detectors is the study of the muon flux variations observed at the ground level associated with diverse transient phenomena. Each detector is assembled on the basis of plastic scintillator slab (150 $\times$ 75 $\times$ 5 cm$^{3}$). Details of experimental setup are reported on \cite{fauth,aug3}.

\subsubsection{The telescope mode}
\label{sec:telescope}

\begin{figure}
\vspace*{-0.0cm}
\hspace*{0.0cm}
\centering
\includegraphics[width=12.0cm]{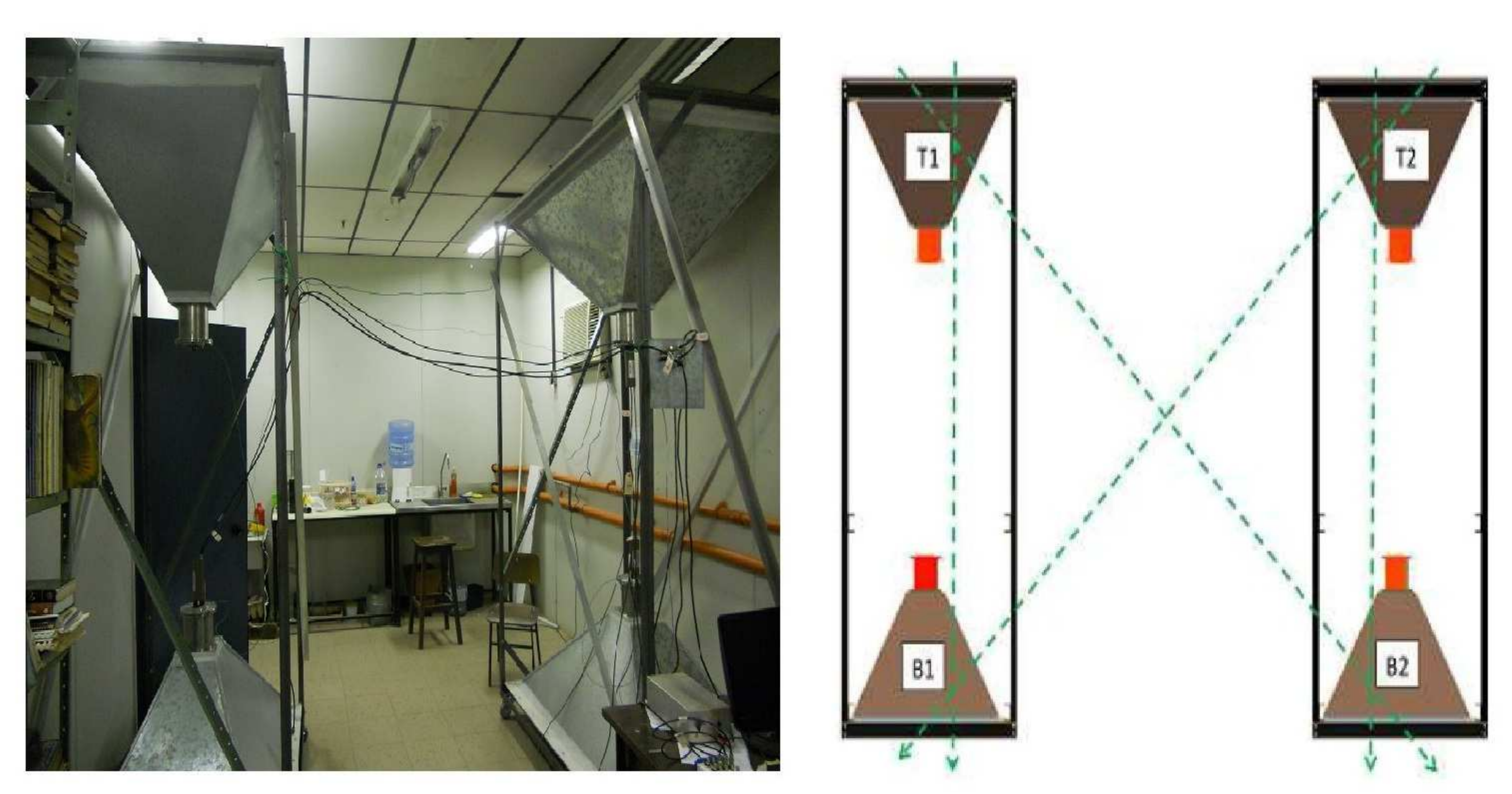}
\vspace*{-0.0cm}
\caption{Left: Photograph of New-Tupi telescope. Right: General scheme of New-Tupi telescope. This configuration allows measuring the muon flux from three directions, the vertical, west, and east. The Muonca telescope has identical configurations.}
\label{fig4}
\vspace*{0.5cm}
\end{figure}

The four detector units are placed in pairs, with T1 (top) and B1 (bottom) and T2 and B2 respectively. This layout allows us to obtain the muon flux from three directions, the vertical, west, and east, the last two with an inclination around 45 degrees. The telescopes record the coincidence rate between T1 and B1; T2 and B2  (vertical), T1 and B2 (west), and T2 and B1 (east). Both vertical and lateral separation between the detectors is 2.83 m. The Fig.~\ref{fig4} summarizes the situation.

\subsubsection{The scaler mode}
\label{sec:scaler}

In parallel with the telescope mode, Muonca is operated in the scaler mode (or single particle technique mode) \citep{obr76,mor84,agl96}, where the single hit rates of all of the four PMTs are recorded once a second. This mode allows us to detect muon flux at fixed time intervals, within a wide field of view. However, the efficiency of particle detection in the scaler mode decreases as the zenith angle of the incident particle increases, due to atmospheric absorption. Thus, the scaler method is limited to incident particles with the zenith angle less than 60 degrees.

\begin{figure}[!h]
\vspace*{-0.0cm}
\centering
 \includegraphics[scale=0.20]{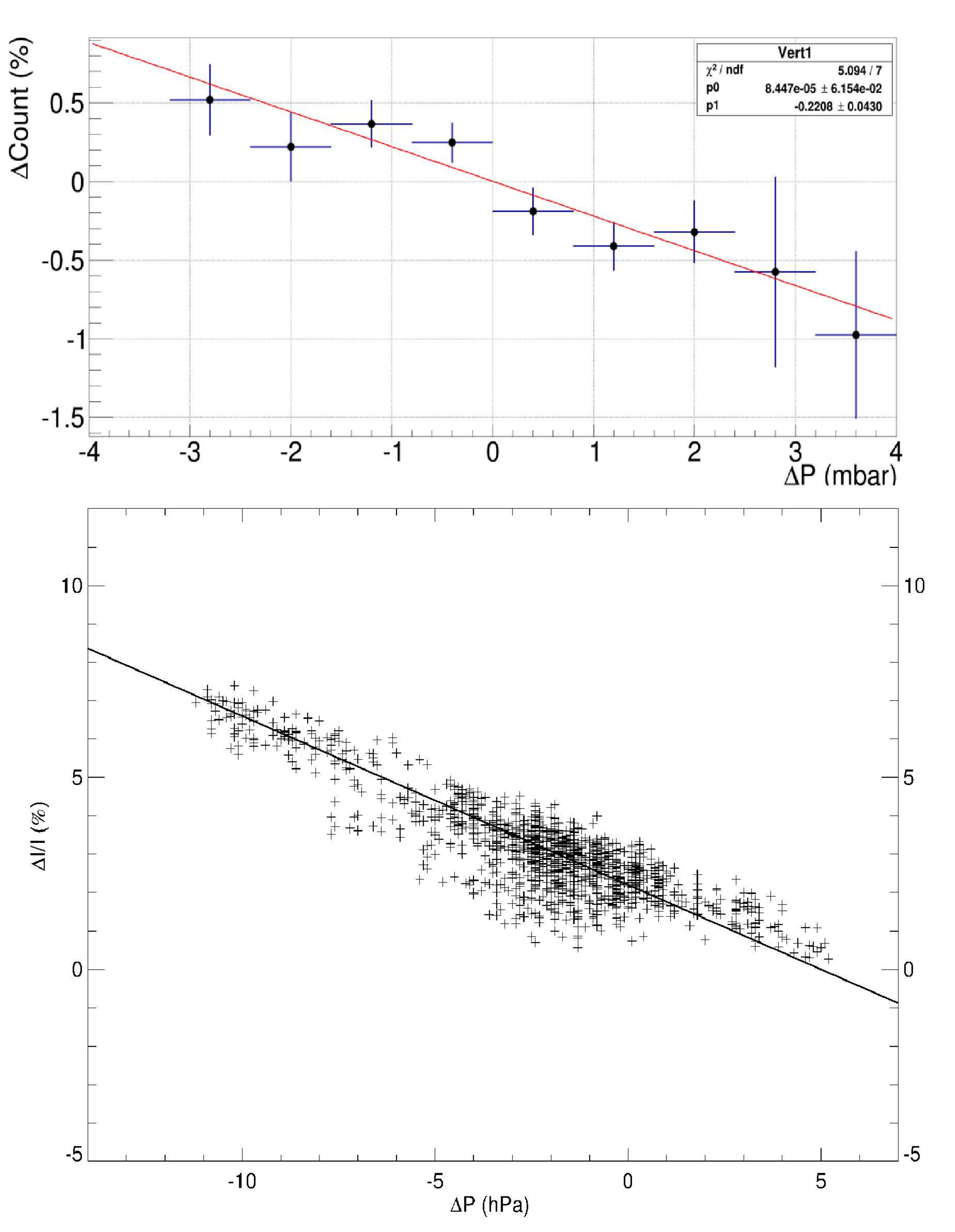}
\vspace*{-0.0cm}
\caption
{
Evaluation of the barometric coefficients. \textit{Top panel}: the muon rate measured by the Muonca telescope versus barometric pressure (in mbar) \citep{fauth}. \textit{Bottom panel}: the relative cosmic ray intensity variation observed by the channel N12 of El Leoncito detector on July 2009 versus the pressure deviation (in hPa) \citep{leo}. The least squared fit estimates give the barometric coefficient $\beta_p = (-0.22\pm 0.04)$ \%/mbar at Muonca \citep{fauth} and $\beta_p = (-0.44 \pm 0.01)$ \%/mbar at El Leoncito \citep{leo}. 1 mbar is equal to 1 hPa.
}
\vspace*{0.7cm}
\label{fig5}
\end{figure}

\subsection{The El Leoncito detector}
\label{sec:leoncito}

The CARPET charged particles detector was developed within an international cooperation between the Lebedev Physical Institute RAS (LPI, Russia), the Centro de Radio Astronomia e Astrofisica Mackenzie (CRAAM, Brazil) and the Complejo Astronomico El Leoncito (CASLEO, San Juan, Argentina). The place of operation of the El Leoncito detector is in the Argentinian Andes. The instrument is based on 240 Geiger type tubes. Details of experimental setup are reported on \cite{leo,leo13b,leo15}.

\subsection{Detector comparisons}
\label{sec:comparisons}

The cosmic ray instruments located at mountain altitudes detect a wide variety of secondary particles, especially neutrons, protons, and electrons. The muon detectors, specifically those located close to the sea level, mostly detect the muon component (often called the hard component, due to their large penetration) of the cosmic rays. However, there are some similarities. For instance, these muon detectors and the cosmic ray neutron monitors have the same energy threshold, close to 100 MeV. Also, for detectors installed at the same location, the corrected counting rate of neutron monitors was confirmed to match the counting rate of the muon detector with a correlation coefficient of 0.8046 \citep{kim12}.

\subsection{Pressure corrections}
\label{sec:pressure}

The cosmic ray intensity as recorded by ground-based detectors varies with atmospheric pressure \citep{dor95,heb02}. The anti-correlated relation between both is well known. In order to remove the atmospheric pressure change influence on cosmic ray intensity observed by a detector, it is necessary to obtain the barometric coefficient, which is inferred experimentally, and in most cases, the value is about $\beta_p \sim - 1.0 \%/mbar$.

The influence of the pressure atmospheric variation on the muon counting rate is weaker. Fig.~\ref{fig5} (top panel) shows the muon rate measured by the Muonca telescope versus barometric pressure (in mbar) \citep{fauth} together with the relative cosmic ray intensity variation observed by the channel N12 of the El Leoncito detector on July 2009 versus the pressure deviation (in hPa) \citep{leo}. The evaluation of the barometric coefficients using the least squared fit estimates gives the barometric coefficient $\beta_p = (-0.22\pm 0.04)$ \%/mbar at Muonca \citep{fauth} and $\beta_p = (-0.44 \pm 0.01)$ \%/mbar at El Leoncito \citep{leo}.

These experimental results are also in agreement with the muon barometric coefficients obtained via Monte Carlo calculations, by means of the CORSIKA code \citep{Kov}. Thus, the barometric coefficient for muons (at sea level or near it) is about ten times (or at least seven times) lower than the barometric coefficient for the neutron monitors in the energy region above 0.1 GeV. 

\subsection{Data sets}
\label{sec:dataset}

In this paper, we examined the most widely used indicators of global magnetospheric activity \citep{may80}, the estimated 3 hr planetary Kp index\footnote{\url{http://www.swpc.noaa.gov/rt\_plots/kp\_3d.html}} and the ring current $Dst$ index\footnote{\url{http://wdc.kugi.kyoto-u.ac.jp/}}. The Kp and $Dst$ indices show different behaviors during different types of solar wind drivers; they also have different sensitivities to the different storm time current systems \citep{wdg94,hut02}. We also used the data measurements of the horizontal component of the geomagnetic field measured by the EMBRACE Magnetometer Network available from \url{http://www2.inpe.br/climaespacial/portal/resumo-de-indices}. 

The EMBRACE magnetometers are scattered in South America and covering the SAA region\footnote{\url{http://www2.inpe.br/climaespacial/portal/variacao-de-h/}} \citep{Dena15}, and the measurements of the horizontal component of the geomagnetic field are done by the difference between the mean of the five most quiet days of the previous month (according to International Q-Days and D-Days, Kyoto) and the H component of each station. The values of the Auroral Electrojet Index (AE) are also utilized in our study. The AE index is the measure of auroral zone magnetic activity produced by enhanced ionospheric currents in the auroral zone in the northern hemisphere \citep{davis66}. We compared our observations with the particle, solar wind and magnetic field available data obtained by spacecraft detectors. 

The Advanced Composition Explorer (ACE) \citep{Stone98} and Solar and Heliospheric Observatory (SOHO) \citep{Har95}. They are in orbit around the Sun, at the  Sun-Earth Lagrange (L1) point (approximately $1.5 \times 10^{6} km$ from Earth) and provide near real-time solar wind information and continuous real-time monitoring of space weather. The Geostationary Operational Environmental Satellites (GOES) circle the Earth in a geosynchronous orbit and provide solar X-ray data, the local magnitude and direction of the Earth's magnetic field, as well as solar proton flux \citep{Rod10}. The Solar Dynamics Observatory (SDO) \citep{pes11} satellite observes the Sun in an inclined geosynchronous orbit and provides heliospheric imagery with the Atmospheric Imaging Assembly (AIA) and Helioseismic and Magnetic Imager (HMI). The number of halo and partial-halo CMEs from the Large Angle and Spectroscopic Coronagraph (LASCO) on-board the SOHO are taken from the CDAW catalog\footnote{\url{http://cdaw.gsfc.nasa.gov}}.

The Computer Aided CME Tracking (CACTus) identification of CMEs is taken from \url{http://sidc.oma.be/cactus/catalog/LASCO/} and analyzed in temporal coincidence with the solar flares\footnote{\url{http://www.lmsal.com/solarsoft/latest\_events\_archive.html}}$^{,}$\footnote{\url{http://www.tesis.lebedev.ru}}. Identification of interplanetary structures and associated solar activity was based on the nomenclature and definitions given by the satellite observations, including an incomplete list of possible interplanetary shocks observed by the CELIAS/MTOF Proton Monitor the SOHO spacecraft\footnote{\url{http://umtof.umd.edu/pm/FIGS.HTML}}. For estimation of the geomagnetic storm onset, we also used the OMNIWeb data services\footnote{\url{http://omniweb.gsfc.nasa.gov}}. For our analysis, we also used data obtained from the High-Resolution Neutron Monitor Database (NMDB) (\url{http://www.nmdb.eu}).


\section{The 2015 Summer Solstice Event}
\label{sec:solstice}

In the middle of June 2015, the Sun's activity increased in one of the biggest sunspot active regions (AR) (NOAA AR 2371)  directly facing Earth. From 18-23 June, the AR 2371 erupted four times. Over the next 5 days there were four M-class flares. On 18 June 2015 active region  2371 displayed a beta-gamma magnetic configuration \citep{hal25}, and was located at N09E50 (solar disc coordinates). It produced a long duration M3.0-class solar flare. The flare started at 16:30 UT, peaked at 17:36 UT and ended at 18:25 UT. A full halo CME was produced in this event. In the CACTus catalog \citep{cac1,cac2} this is CME 0078. It was observed by SOHO at 17:24 UTC on 18 June. A partial halo CME seen by SOHO/LASCO C2  was associated with a large solar filament eruption in the SSE quadrant (S27E06) of the solar disc observed in SDO/AIA imagery on 19 June. AR 2371 with the changed magnetic configuration (beta-gamma-delta) and produced a pair of  M2-class flares (located at N12E13 and N12E16) on 21 June. Also associated with this flare, CME 0090  became visible in LASCO C2 imagery at 02:36 UT. Also within the time period of the dual flare, CME 0091 occurred at 02:48 with strong shock waves propagated in the western region of the ecliptic plane at $\sim 270$ degrees. On 22 June the same active region with a beta-gamma configuration (located at N13W06) produced an M6.5 solar flare at 18:23 UT. A full halo CME, 0093, was produced and became visible in LASCO C2 imagery at 18:36 UT.

\begin{figure}
\vspace*{0.0cm}
\hspace*{-1.0cm}
\centering
\includegraphics[width=9.0cm]{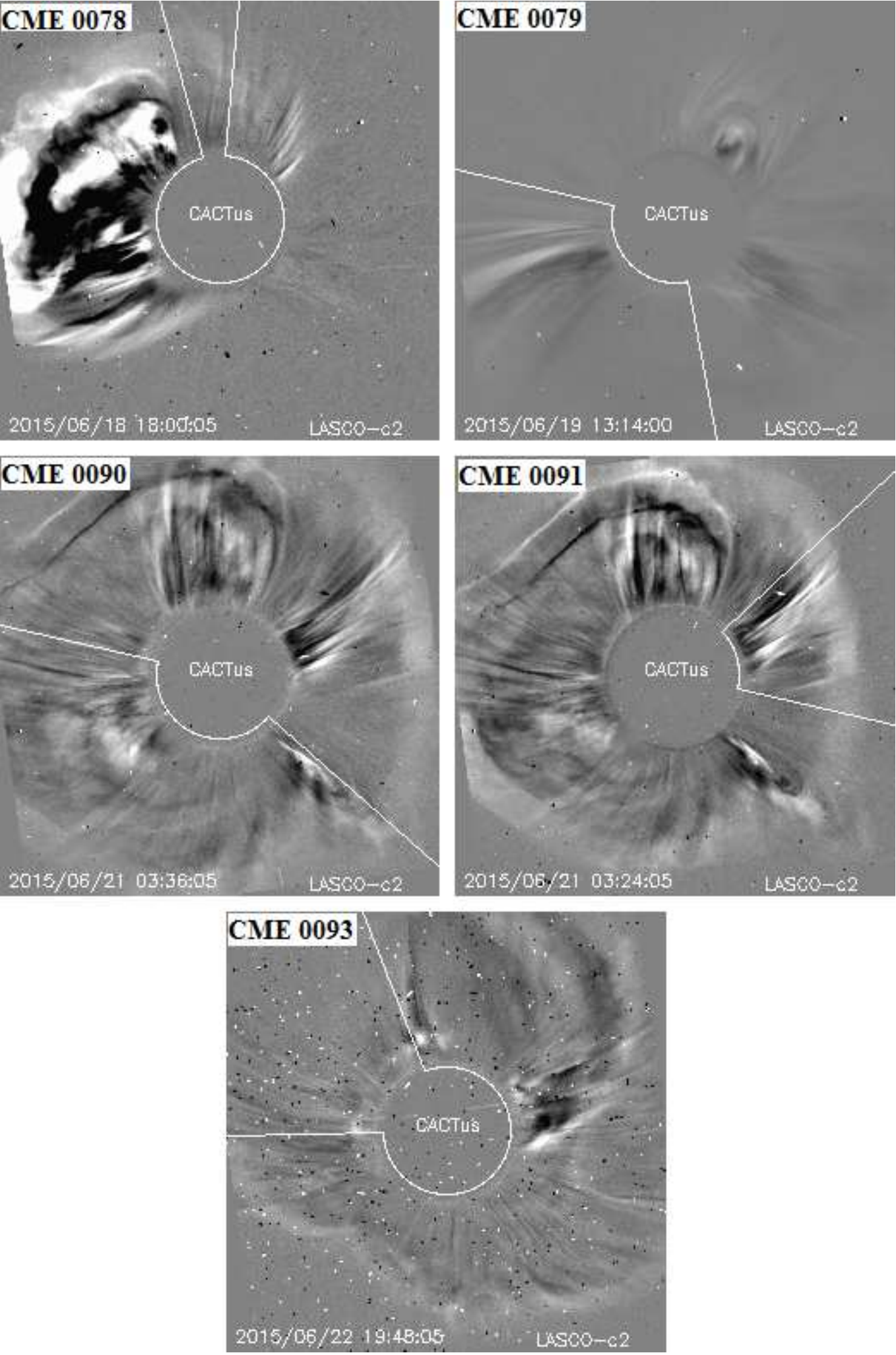}
\vspace*{0.0cm}
\caption
{
Five CMEs were detected in the period 18-22 June 2015. The full halo CME 0078 is associated with an M3-class flare on 18 June. The partial halo CME 0079 is associated with the filament eruption in the S27E06 region on 19 June. CME 0090 and CME 0091 are time correlated with the double-peaked M2-class flare of 21 June. CME 0093 was produced by a long duration M6.5 solar flare on 22 June. Four CMEs were associated with flares in the active region AR 2371. Credit: SOHO LASCO C2 and the CACTus catalog.
}
\label{fig6}
\end{figure}

\begin{figure}
\hspace*{0.0cm}
\centering
\includegraphics[width=13.0cm]{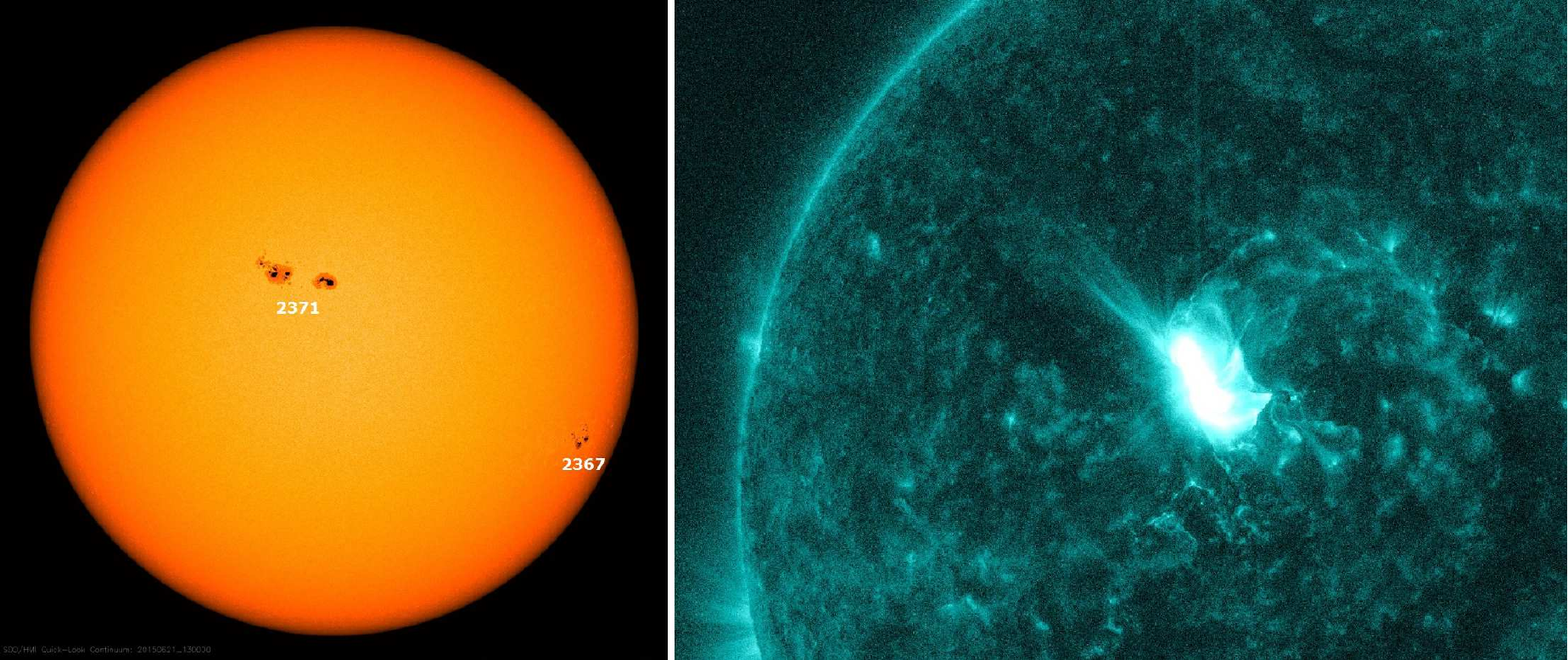}
\caption
{
Left: Sunspots on 21 June 2015, highlighting an active region AR 2371. It was one of the largest sunspots in the current solar cycle 24. 
Right: Solar blast surrounding an active region AR 2371 on 21 June 2015. It was the origin of an Earth-directed CME. Credit: SDO/AIA/HMI.}
\label{fig7}
\end{figure}

The images of CME 0078, 0079, 0090, 0091, 0093 are presented in  Fig.~\ref{fig6}. The flare site is shown in Fig.~\ref{fig7} indicating the active region AR 2371.

\subsection{The first CME impact on Earth's magnotosphere}
\label{sec:fimpact}

According to the SOHO CELIAS/MTOF Proton Monitor on the SOHO spacecraft, in the period 21-22 June, three interplanetary shocks passed the L1 point. The first one was detected on 21 June at 15:50 UTC. The weak impact with the Earth's magnetic field was seen at $\sim$ 16:45 UT by Boulder USGS Magnetometer\footnote{\url{https://geomag.usgs.gov/storm/storm23.php}}. The source of this interplanetary disturbance is estimated to be the eruption of the CME 0078 on 18 June. The travel time of the CME 0078 to reach Earth was $\sim$ 71 h withing the expected for a CME with an average speed of 525 km s$^{-1}$. However, his impact did not cause an immediate geomagnetic storm.

\subsection{Second and third impacts on Earth}
\label{sec:stimpact}

\begin{figure}
\vspace*{0.0cm}
\hspace*{-0.0cm}
\centering
\includegraphics[width=12.0cm]{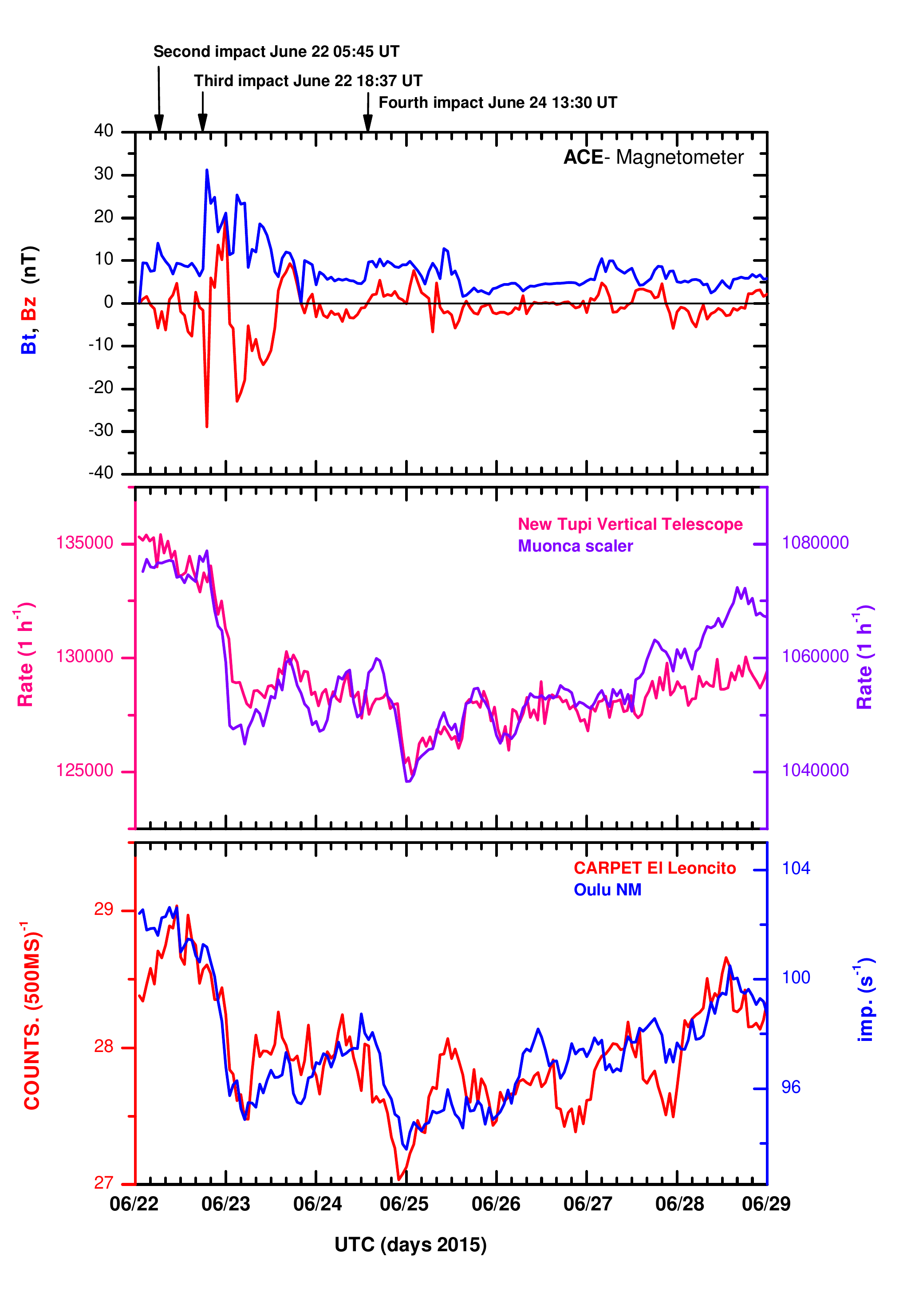}
\vspace*{-0.5cm}
\caption
{Comparison between the IMF strength (Bt) and the IMF Bz component observed by the ACE satellite (\textit{top panel}) and the 1-hour muon counting rate in Muonca (scaler data)  and New-Tupi vertical telescope (\textit{bottom panel}) in the period 22-23 June 2015. The vertical arrows show the onset time at Earth of the interplanetary shocks originated by the CMEs.}
\label{fig8}
\end{figure}

\begin{figure}
\vspace*{0.0cm}
\hspace*{-0.0cm}
\centering
\includegraphics[width=12.0cm]{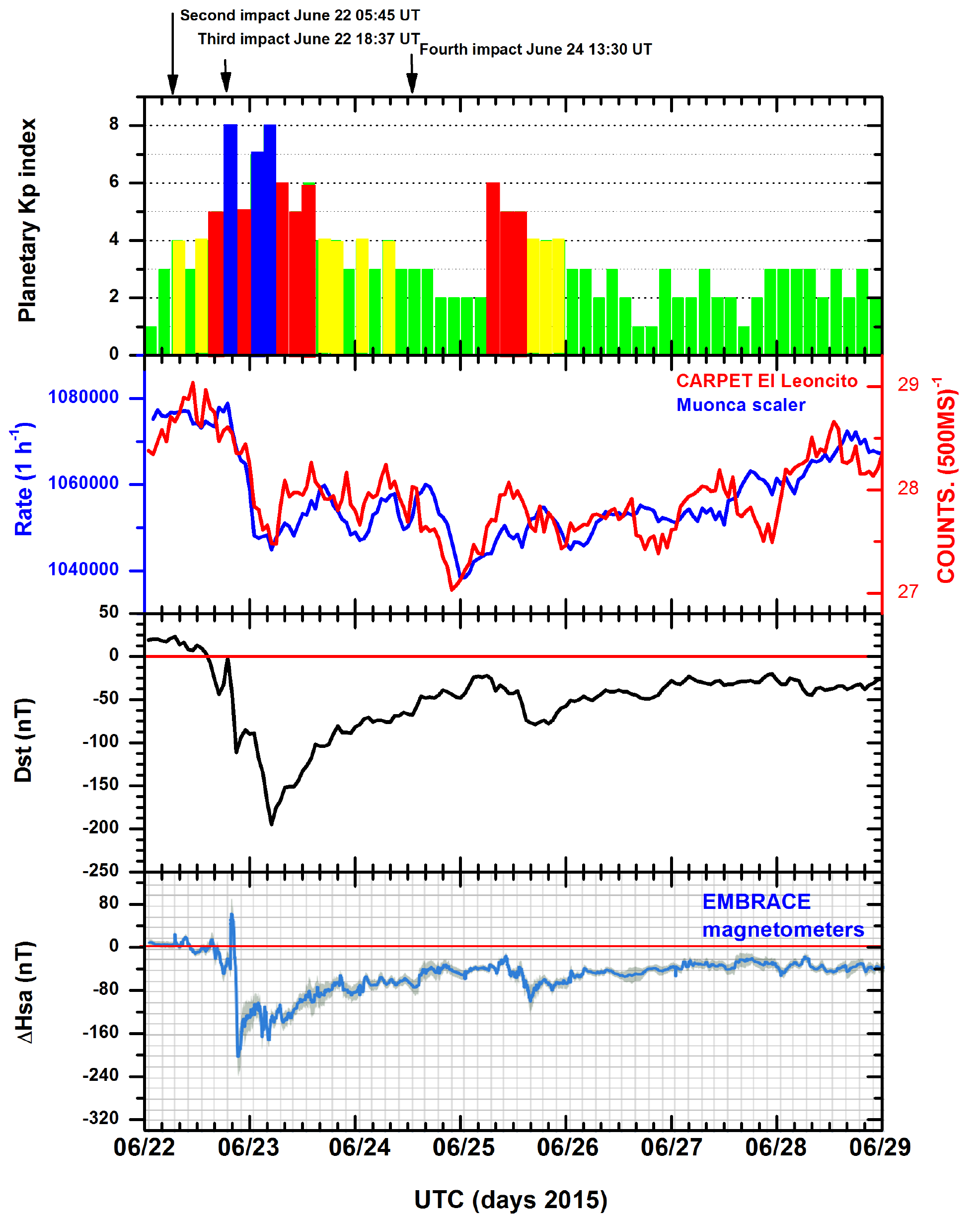}
\vspace*{-0.0cm}
\caption
{From top to bottom, the panels show the Kp index (\textit{top panel}), the 1 hour binning muon counting rate in Muonca (scaler data) and cosmic ray counting rate in the detector CARPET El Leoncito (\textit{2nd panel}), the $Dst$ index (\textit{3rd panel}) and the EMBRACE magnetometers (\textit{4th panel}) in the period 22-23 June 2015. The black arrows on the top of the figure indicate the onset time at Earth of the interplanetary shocks originated by the CMEs.}
\label{fig9}
\end{figure}

On 22 June at 04:52 UT, a second interplanetary shock crossed L1 and the shock arrived at Earth at 05:45 UT. There were subsequent enhancements in solar wind velocity and density, i.e., a third interplanetary shock crossed L1 at 18:01 UT. These impacts were classified as shock waves by SOHO CELIAS/MTOF\footnote{\url{http://umtof.umd.edu/pm/FIGS.HTML}}. The third impact at Earth was registered at $\sim$ 18:37 UT, and only after this third impact, the ground-level detectors observed depressions of galactic cosmic ray intensity, the FD. The superposition effects of the succession of passing ICMEs and the impacts in the Earth's magnetic field and surrounding regions (such as the L1  triggered the 22-23 June (G4 severe-class) geomagnetic storm ($Kp = 8$). This strong perturbation of the geomagnetic field was the so-called 2015 Summer Solstice storm. Fig.~\ref{fig8} and Fig.~\ref{fig9} summarize the situation. 

In section~\ref{sec:check} we will discuss a cross-check between the CACTus CME catalog parameters, the associated geomagnetic parameters and a possible association between the second and third impacts.

Fig.~\ref{fig8} shows the comparison between the Bt and Bz solar wind magnetic component observed by the ACE satellite at L1 (top panel), the 1 hour muon counting rate in New-Tupi vertical telescope, the Muonca scaler data (central panel), the counting rate in the El Leoncito detector and the Oulu NM data (bottom panel). The black arrows at the top of the figure show the onset time at Earth of the three interplanetary shocks originated by the CMEs. As can be seen from Fig.~\ref{fig8}, the Bz magnetic component made oscillations in coincidence with the second impact. A prolonged period of mostly southward Bz occurred after the third impact and remained negative from $\sim$ 01:50 UT until $\sim$ 06:00 UT on 23 June, reaching as low as $\sim\;-25$ nT. The FD intensity (depression) reached an amplitude of up to 4\% in New-Tupi vertical telescope and Muonca. In addition, note a good agreement between New-Tupi vertical telescope and Muonca scaler data.

Fig.~\ref{fig9} shows a good temporal correlation between the Kp index (top panel), 1 hour binning muon counting rate in Muonca (scaler data) and the cosmic ray counting rate corrected for pressure and temperature effects in the detector at El Leoncito (second panel), the $Dst$ index (third panel) and the EMBRACE magnetometer data in the period 22-23 June 2015. The planetary geomagnetic index Kp shows a period of enhanced geomagnetic activity reaching the level from G3 ($Kp=7$) to G4 ($Kp=8$) at $\sim$ 18:50 UT on 22 June, with the arrival of the third shock, as shown in Fig.~\ref{fig8} (top panel).

The El Leoncito cosmic ray detector measured the FD amplitude of about 3.7\%. One can see that the maximum intensity of the FD and the $Dst$ index of about $-200$ nT are almost in temporal coincidence due to the third impact on 22 June, both reaching the minimum value at the early hours on 23 June. The third impact has a good temporal correlation with data from the EMBRACE magnetometers that measured $-198$ nT for the horizontal component of the geomagnetic field at $\sim$ 20:30 UT on 22 June.

The AE index showed strong variations 2 h after the second impact with a peak above 1,000 nT at 07:30 UT\footnote{\url{wdc.kugi.kyoto-u.ac.jp/ae\_provisional/201506/index\_20150622.html}}$^{,}$\footnote{\url{wdc.kugi.kyoto-u.ac.jp/ae\_provisional/201506/index\_20150623.html}}. The third impact showed a temporal correlation with variations in the AE index exceeding 2,000 nT in the time period between 18:30 and 20:30 UT. The AE index followed an oscillatory behavior, often reaching values above 1,000 nT until $\sim$ 13:00 UT on 23 June. It was a severe geomagnetic storm, one of the strongest in the current solar cycle 24.

\subsection{The fourth impact on Earth}
\label{sec:foimpact}

The fourth impact was observed at 13:30 UT on 24 June with the arrival of a shock wave from CME 0093 on 22 June. It is indicated by the last vertical arrow in Fig.~\ref{fig8} and Fig.~\ref{fig9}. A small fall in the muon counting rate was observed by New-Tupi, Muonca, CARPET, and Oulu. Fig.~\ref{fig7} and Fig.~\ref{fig8} summarize the situation. This is an example of a mini-FD event \citep{nav05}. Usually, they indicate an interplanetary disturbance that crosses the vicinity of Earth. However, in the present case, the situation is different, since the origin of the FD was a full halo CME directed towards Earth (see section~\ref{sec:check}).

\begin{figure}
\vspace*{-0.0cm}
\hspace*{0.0cm}
\centering
\includegraphics[width=12.0cm]{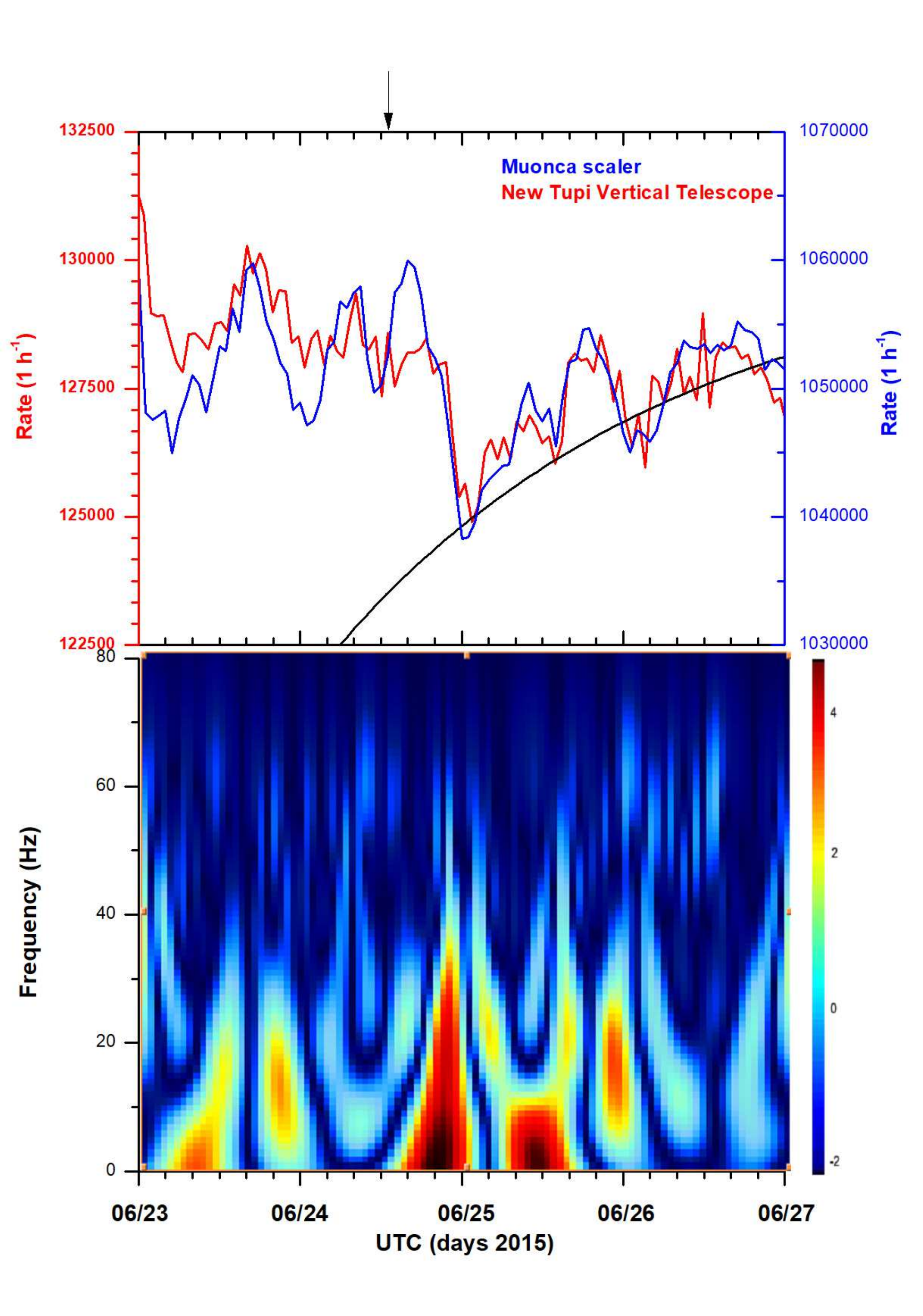}
\vspace*{-1.0cm}
\caption{
\textit{Top panel}: the counting rate observed by New-Tupi muon telescope and Muonca (scaler data) on 23-26 June 2015. To estimate the recovery time of mini-FDs we use a nonlinear fit of the data to a single exponent (black curve). \textit{Bottom panel}: A time-frequency representation of the combined (New-Tupi telescopes and MUONCA) data for the same period using the continuous wavelet transform (CWT) analysis. The vertical color bar indicates the relative power (variance). High power is indicated by warm (red) colors, whereas low power is indicated by cold (blue) colors. The intervals corresponding to the fall (24-25 June 2015) and the mini-FDs (25-26 June 2015) observed in the counting rate can be easily identified. The black arrows on the top of the figure correspond to CME 0093.}
\label{fig10}
\end{figure} 
 
As can be seen in Fig.~\ref{fig9} (top panel), the effect of this fourth CME impact on the planetary Kp index became evident in early hours on 25 June, producing a moderate geomagnetic storm of level G2 ($Kp=6$). There is also a small signal in the $Dst$ index and the EMBRACE magnetometers data shown in Fig.~\ref{fig9} (third and fourth panels). 

We note that the time of impact of the interplanetary disturbance attributed to CME 0093 (last CME) is in temporal coincidence with the onset of the mini-FDs (see Fig.~\ref{fig8} and  Fig.~\ref{fig9}), whereas there is a delay ($\sim 16$ h) between the impact of the CME and the onset of the geomagnetic storm reflected by the Kp and $Dst$ indices. In addition, there are no signs of significant disturbances in the AE measurements in temporal coincidence with the fourth impact on 24 June \footnote{\url{wdc.kugi.kyoto-u.ac.jp/ae\_provisional/201506/index\_20150624.html}}. The AE index again showed values above 1,000 nT at 7:30 UT on 25 June\footnote{\url{wdc.kugi.kyoto-u.ac.jp/ae\_provisional/201506/index\_20150625.html}}, showing similar dynamic behavior with the Kp and $Dst$ indices.

We would like to point out, that the fourth impact occurred under very special conditions, i.e., it was already perturbed geomagnetic condition
due to two previous impacts.

In order to investigate the effect at the ground level of the fourth impact (CME 0093) on 24 June, we analyzed data from New-Tupi and Muonca (scaler data) using the continuous wavelet transform (CWT) \citep{tor98,ryb01}. The Fourier spectrum of a signal represents the frequency content of the signal, the signal itself is in the time domain. The frequency spectrum in time-frequency maps is given for each time step so that one can see the evolution of the frequencies. The CWT spectra were obtained for the combined data between New-Tupi and Muonca (scaler data) and are presented in Fig.~\ref{fig10} (bottom panel). In this case, both signals show high power variance. High power (variance) is indicated by red color whereas low power is indicated by blue color. One can see that the mini-FDs (top panel Fig.~\ref{fig10}) is correlated with the narrow peak of the CWT, both are identified. 

The forth impact (originated by CME 0093) has stretched the total duration (from the onset time until the recovery time) of the main summer solstice FD for at least 48 h, as shown in Fig.~\ref{fig10} (top panel). The estimated mini-FDs recovery time interval is defined as the time required for the counting rate to return from the maximum depression to the pre-FDs level. According to \cite{uso08}, the shape of the recovery phase of an FD can be approximated by an exponential recovery function $I = I_0 \times A \exp(-t/\tau)$, where $I$ and $I_0$ are the current and undisturbed CR intensities, A is an amplitude and $\tau$ is the characteristic recovery time. In the present case, the situation is more complex, because the mini-FDs occurred in a previously perturbed geomagnetic field. However, it still can be described by an exponential function. The mini-FDs recovery time is estimated as, $\tau = (51 \pm 5)$h. The bold line on Fig.~\ref{fig10} (top panel) summarizes the situation. Finally, the recovery time of the Summer Solstice FD is estimated as 144 h.


\section{Cross-check between the CACTus CME catalog, FD events and the geomagnetic parameters}
\label{sec:check}

In this section we discuss the CACTus CME parameters and compare them with the geomagnetic parameters, as well as experimental data on FDs.

\subsection{The CACTus catalog}
\label{sec:thecactus}

The images of the CMEs derived from the coronagraph observation (SOHO-LASCO C2) have only an apparent geometric meaning, since they depend on the orientation of the CME in the relation of the observer (L1). The outputs given by CACTus include information on the CME identification number, onset time (earliest indication of liftoff), position angle (pa) (counterclockwise from solar north; degrees), angular width (da) in degrees, median velocity (km s$^{-1}$), and the highest velocity detected within the CME
\citep{cac1,cac2}.

The angular width derived from projected images is only an apparent quantity, which indicates the angular size of the CME volume projected onto the plane of the sky. The $da$ distribution is essentially scale-invariant, which means that there is no typical size of a CME. Most CMEs have an angular width around 30$^{\circ}$, and the number of CMEs decreases as the angular width increases, following a power law distribution. If $da > 90^{\circ}$, then a CME is classified as type II, while a CME with 
$da > 180^{\circ}$ is called type III or partial halo. Halo CMEs are those with $da > 270^{\circ}$ (or type IV). A CME appears as a halo or partial halo if it is launched in a path close to the Sun-Earth direction. This explains, at least in part, why most geomagnetic storms triggering FD are associated with full halo CMEs (e.g., \cite{belov14}).

The position angle $pa$ is correlated with the CME projected latitude, and it is defined as the middle angle of the CME when seen in white-light images and depends on the orientation of the CME in the relation of the observer (L1). Thus, the projected latitudes are only an approximation of the true direction of propagation. Values of the $pa$ close to 90 and 270 degrees represent zero latitudes. When $pa = 90^{\circ}$ and $pa = 270^{\circ}$, this means that the middle angle of the CME coincides with the eastern side and with the western side of the ecliptic plane, respectively. We will show that CMEs with a narrow width angle can trigger geomagnetic storms only if their $pa$ is close to 270 degrees, i.e., close to the western region of the ecliptic plane. The CACTus catalog shows that the latitude dependence of the CMEs is approximately a flat (random) distribution, with some structure; it is strongly restricted to two broad bands around $\sim 50^{\circ}$ latitude, and also depends strongly on the solar cycle time. 

\begin{figure}
\vspace*{-0.0cm}
\hspace*{0.0cm}
\centering
\includegraphics[width=10.0cm]{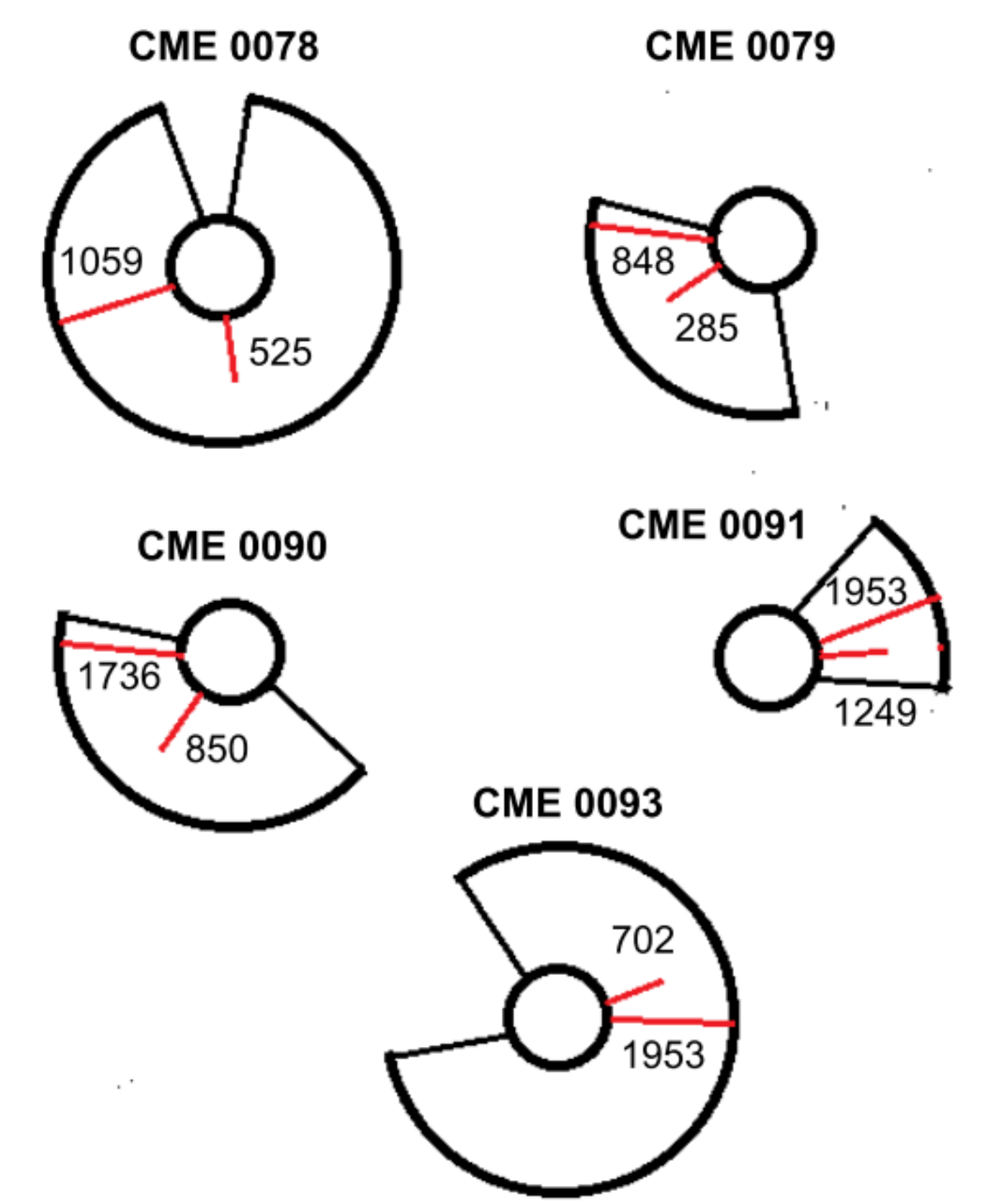}
\vspace*{-0.0cm}
\caption
{
The CACTus parameters (position angle, angular width, median and maximum speed) of the five CMEs registered in the period 18-22 June 2015.
The median CME speed is oriented according to the pa. The images of the CMEs observed by SOHO LASCO C2 are presented in Fig.~\ref{fig6}.
}
\label{fig11}
\end{figure}

The speed profile measured by CACTus shows that CMEs have internal speed variability of the plasma structure. This behavior is probably due to interaction with different background solar wind structures. CACTus measures a linear speed profile as a function of the $pa$ (latitude) and lists the median and maximum speed values, while the previous catalog, such as the CDAW\footnote{\url{http://cdaw.gsfc.nasa.gov/CME\_list}} \citep{yashiro04, gop09} reported only the fastest moving feature of the leading edge. The CACTus CME speed is log-normally distributed, peaking at 200-400 km s$^{-1}$. The uncertainty in the speed is larger during the solar maximum and is higher than 175 km s$^{-1}$. Fig.~\ref{fig11} summarizes the CACTus parameters of the five CMEs detected in the period 18-23 June 2015. The external contour represents the angular width, while the short and long red lines within each area show the median and the maximum speed, respectively. The median CME speed is oriented according to the $pa$.

The width above 270$^{\circ}$ indicates that CME 0078 and CME 0093 are halo CMEs. We can see from Fig.~\ref{fig11} that CME 0079 of 19 June (the solar filament eruption) was directed to the southeast region, below the ecliptic plane. There is no signal of the passage of the CME ejecta associated with this blast by L1 and, consequently, there was no significant impact on the magnetosphere is expected from this eruption. Thus, it is likely that CME 0079 was not associated with the second impact on Earth on 22 June (Fig.~\ref{fig8} and Fig.~\ref{fig9}).

\subsection{Cross-check between the CACTus catalog and ground level observations}
\label{sec:foimpact}

The next step is to cross-check the CACTus catalog and the ground-level observations. We are looking for the particular features of the CMEs that triggered geomagnetic storms and were associated with the observed FD events. Previous studies \citep{gop09} show that a typical geoeffective CME is characterized by (a) high speed, (b) a large angular width (mostly halos and partial halos), and (c) a solar source location close to the central meridian. In general, a full halo CME means a CME directed to the Earth and a narrow CME with $pa\;270^{\circ}$ means a CME propagating in the ecliptic plane and only its projection in the coronograph points to the west. Thus if a full CME and a narrow width CME with  $pa\;270^{\circ}$ originate around the same time within the same active region and with approximately the same speed, then they have almost the same travel time. This is approximately to the case of CME 0090 (partial halo) and the CME 0091 (narrow width). These CMEs erupted with high velocities and almost simultaneously from the same active region AR 2371, when it was close to the central meridian (see Fig.~\ref{fig7}). CME 0090 erupted almost in the Sun-Earth direction (partial halo event), with the pa to the solar southeast. Despite the narrow width angle, the $pa$ of CME 0091 was close to 270$^{\circ}$, the bulk of the CME material was clearly propagating in the westward direction. Thus, the second and third impacts of 22 June were likely associated with the CME 0090 and the CME 0091, respectively. From Fig.~\ref{fig11} we can see that the narrow CME 0091 and the full halo CME 0093 had similar pas close to 270$^{\circ}$. 

In addition, it was shown above that only after the impact of CME 0091 on 22 June (third impact) the G4 geomagnetic storm was triggered, and only after the impact of CME 0093 on 23 June (fourth impact) the G2 geomagnetic storm was triggered. In this interpretation, the portion of CME 0090 which encountered Earth was very weak. Most of the plasma was ejected to the southeast direction as can be seen in Fig.~\ref{fig11}. Hence, the second impact of 22 June is consistent with the CME 0090 origin.

\subsection{Comparison with other FD events}
\label{sec:foimpact}

Here  we perform a cross-check between the CACTus CME catalog, the ring current $Dst$ and  the FD events listed in \cite{lin16}. These events are from the rising phase (2010-2013) of solar cycle 24. Basic information on the selected FD, the CME parameters from the CACTus CME catalog and the ring current $Dst$ indices are shown in Table 2 of the appendix. The FD parameters are based on the Oulu NM data, the FD data listed in \cite{lin16} and New-Tupi telescope data. 

\begin{figure}
\vspace*{-0.0cm}
\hspace*{0.0cm}
\centering
\includegraphics[width=12.0cm]{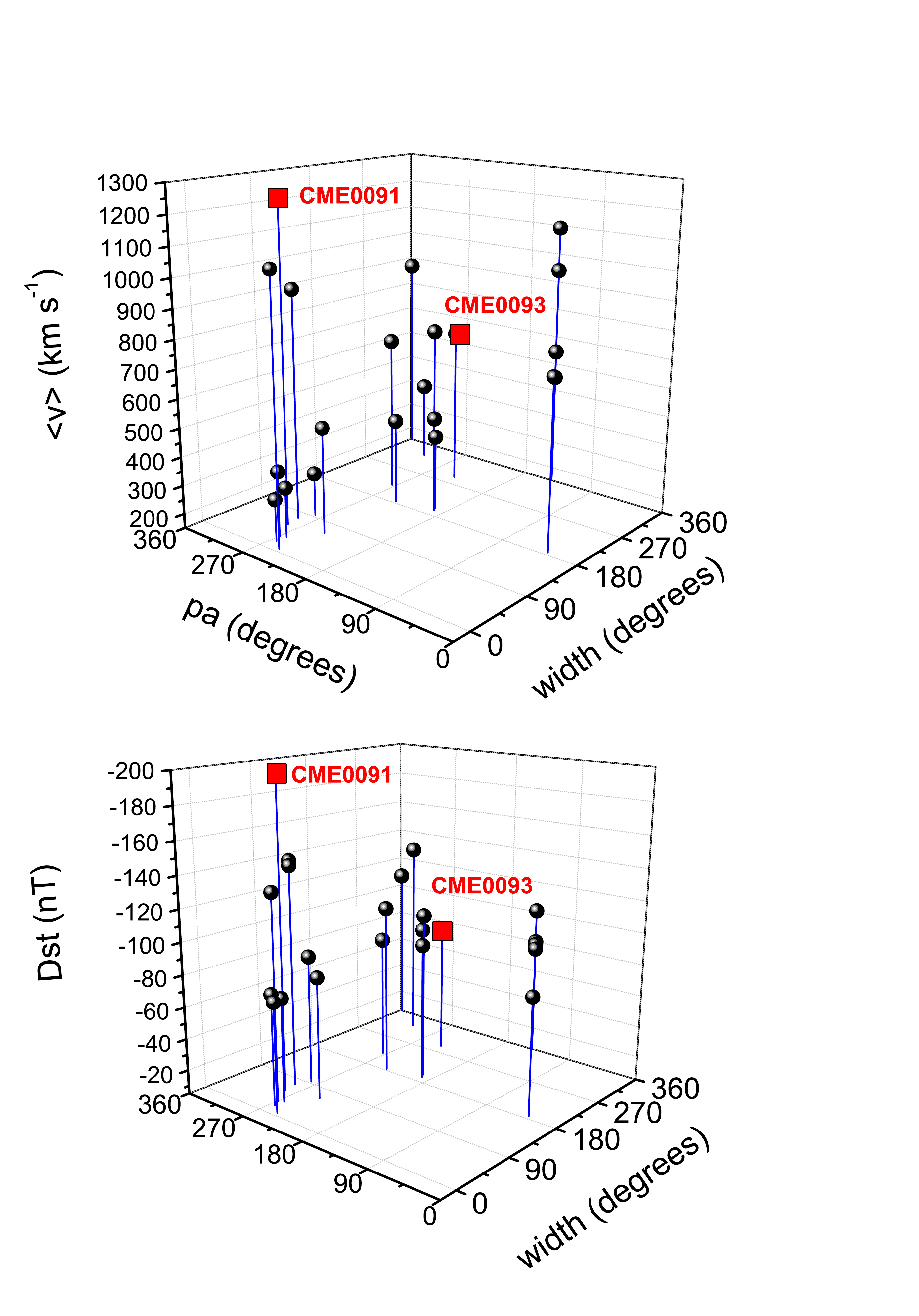}
\vspace*{-0.5cm}
\caption
{The relationship between the CME parameters $pa$, $da$, and the median CME speed (\textit{top panel}). In the \textit{bottom panel} the CME speed was replaced by the $Dst$ index. In both cases, only events with a $Dst$ index less than $-70 nT$ are considered. Squares indicate the solstice storm data.
}
\label{fig12}
\end{figure}

Initially, we have selected only the CMEs that are estimated to have triggered intense geomagnetic storms with a $Dst$ below $-70$ nT. 
These results are shown in Fig.~\ref{fig12}.

In Fig.~\ref{fig12} we can recognize two basic groups: (a) The halo and partial halo events, those with $da>180^{\circ}$, in this case, the $pa$ of the events are distributed within a broad range, i.e., from 0 to 360 degrees. (b) The narrow-angle width events, those with $da<90^{\circ}$. However, in this case, the $pa$ of the events are restricted to a narrow region, mainly around $pa \sim 270^{\circ}$, i.e., on or close to the ecliptic plane.

\begin{figure}
\vspace*{-0.0cm}
\hspace*{-1.5cm}
\centering
\includegraphics[width=13.0cm]{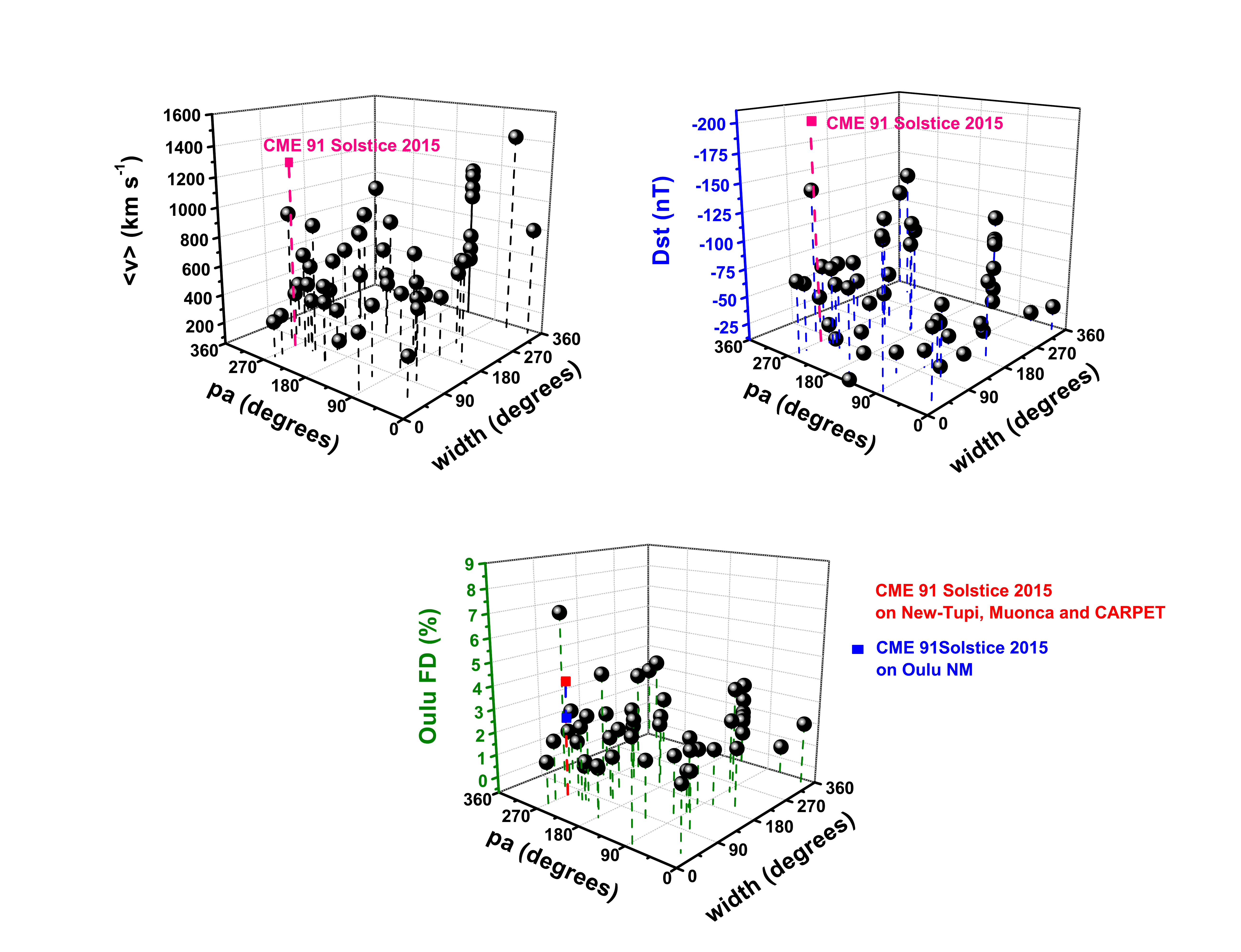}
\vspace*{-0.0cm}
\caption
{The relationship between the CME parameters $pa$, $da$, and the median CME speed (\textit{top left panel}). In the \textit{top right panel} the median CME speed is remplaced by the $Dst$ index. In the \textit{bottom panel} the median CME speed is replaced by the amplitude of the  FD based on Oulu NM data. In all cases, only events with an FD with amplitude equal or above 1\% are considered. Squares indicate the solstice storm data, in this case the FD data are from New-Tupi detector.}
\label{fig13}
\end{figure}  

On the other hand, Fig.~\ref{fig13} shows all the CMEs that triggered FD events with an amplitude equal or above 1\% (at Oulu NM) and for the same time period. Despite a dispersion higher than in the previous case, the CMEs with a narrow width angle are primarily clustered at the $pa$ around $270^{\circ}$. This is the case of the 21 June CME 0091 associated with one of the strongest geomagnetic storms of current solar cycle 24. 

\section{Solar energetic particles and solstice storm 2015}
\label{sec:radiation}

Solar Energetic Particle (SEP) events are also known as radiation storms, or solar proton events (SPE). Solar radiation storms typically occur after major eruptions on the Sun because of charged particles accelerating in the solar atmosphere to very high velocities. Two classes of solar proton events can be distinguished. The first is a tiny fraction of SEP, those with higher velocities, whose acceleration is during the short (typically less than 20 minutes) impulsive phase of the flare. However, the majority of SEP events are associated with solar protons accelerated by CME shocks during the gradual phase of the flare \citep{augJap}. The characteristic of the gradual phase is a soft rise and a long duration decay time of the soft X-ray component. 

A SEP with energies above 100 MeV can trigger so-called ground-level events (GLEs). GLEs are observed as particle excess in the counting rate of ground-level detectors. Three or more stations must observe an increase of 4\% or more in a 3-min moving average \citep{kuwa06,sovo09} as the criteria for the classification of an event as a GLE.

The observation of SEP events implies the existence of coronal shocks extending at least $\sim 300^{\circ}$ \citep{cli95} and interplanetary shocks up to $\sim 180^{\circ}$ at 1 AU \citep{can96}. Neither of the two consecutive CMEs that erupted on 21 June 2015 and were linked to a double M class flare on 21 June, responsible for triggering the severe geomagnetic storm on 22 June, satisfied these criteria. We looked for a possible GLE signal in the NMDB and other ground-based detectors, including New-Tupi and Muonca detectors, but did not find the signal.

According to the ACE-SIS and GOES data, there was a radiation storm with onset at 21 June, associated with the M-class dual flare. The radiation storm reached the S1 (minor) level at 19:36 UT at L1, and one hour later, at 20:38 UT, the S1 level was registered by the (geostationary satellite) GOES. The S1 level indicates that the flux of protons with energies above 10 MeV equals or exceeds 10 $pfu$ (1 $pfu = 1$ particle per $cm^2\;s\;sr$). It is also worth noting that a sudden increase in the speed of the solar wind, from 400 km $s^{-1}$ to almost 700 km $s^{-1}$, was observed due to the likely passage of the CME 0091 ejecta at the L1. This is exactly the signature of the presence of shock waves coinciding with the magnetic disturbance, accelerating particles of the interplanetary medium and even particles from the radiation storm itself.

\begin{figure}
\vspace*{-0.0cm}
\hspace*{-1.0cm}
\centering
\includegraphics[width=12.0cm]{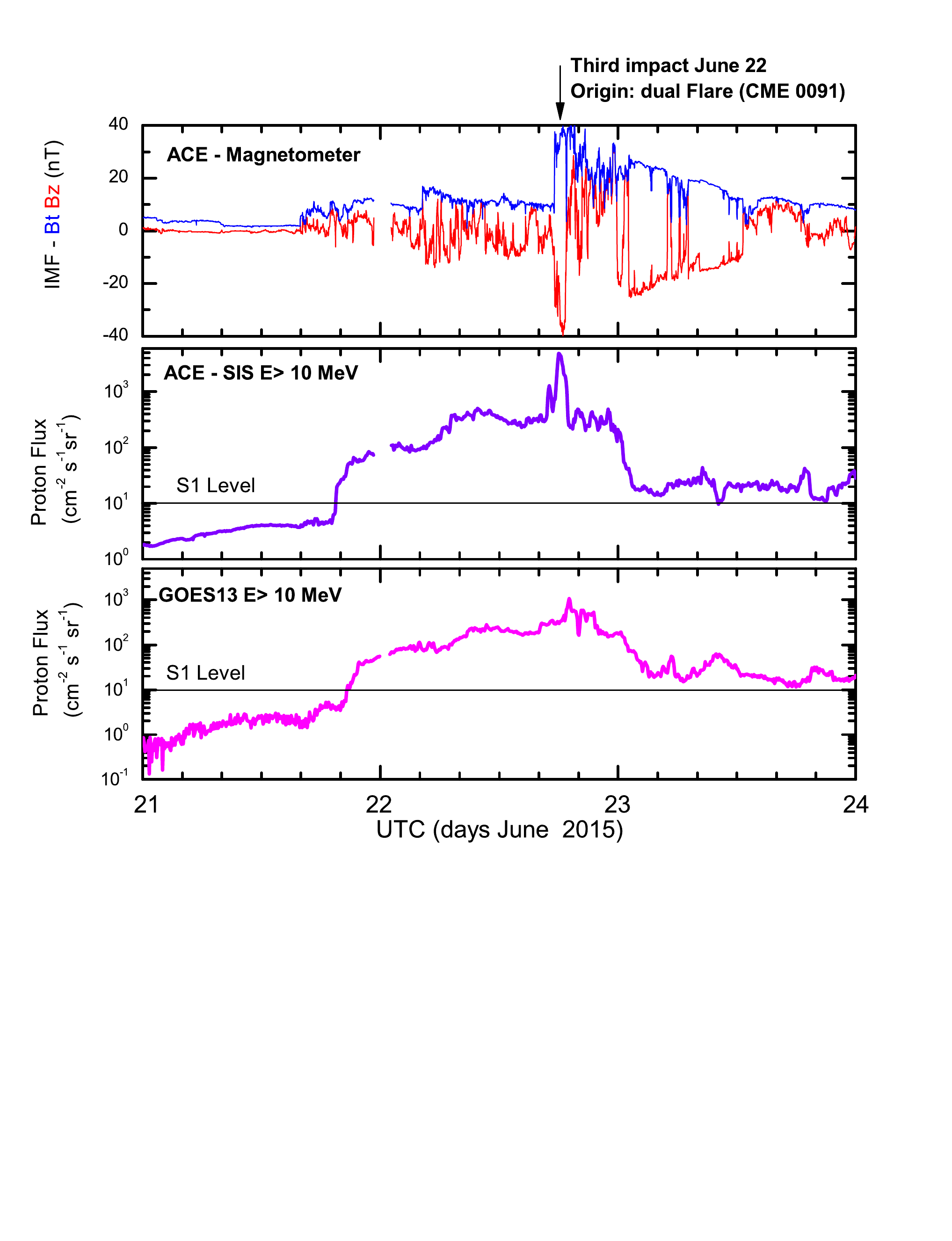}
\vspace*{-5.0cm}
\caption
{Comparison between the magnetic field Bt and Bz components observed by the ACE satellite (\textit{top panel}), the integral protons flux above 10 MeV from Solar Isotope Spectrometer (ACE-SIS) and the energetic (10 MeV) proton flux time profile observed by the GOES 13 satellite (\textit{bottom panel}) in the period 21-23 June 2015. The horizontal line indicates the radiation storm S1 level. The arrow at the top shows the time of the third impact on 22 June.}
\label{fig14}
\end{figure} 

A high-speed magnetized plasma and the Earth's magnetic field oriented in the opposite direction can interconnect through a process that is known as ``magnetic reconnection'' \citep{dore08}. The reconnection process enhances the direct magnetic connection between the solar wind and Earth's magnetosphere. On June 22, this process occurred due to strong southward IMF of the CME 0091 with the geomagnetic field (see top panel on Fig.~\ref{fig14}).

An additional increase in the particle flow can lead to a situation when a more significant fraction of cosmic rays is injected into the upper atmosphere reaching up to middle latitudes and producing an increased rate of particles at ground level. Recent it was reported by GRAPES Collaboration \citep{moh16} that the geomagnetic field perturbed via reconnection with the IMF could lead to a sudden lowering of the cutoff rigidities and an enhancement of the galactic cosmic rays intensity. 

The GRAPES Collaboration has reported a short duration bundle of muons in temporal coincidence with the third impact on 22 June 2015. In addition, there was an increase in the radiation storm level, from S2 (moderate-level) to S3 (strong-level) in the NOAA storm scale. Fig.~\ref{fig14} summarizes the situation.

\section{Summary and conclusions}
\label{sec:conclusion}

One of the strongest geomagnetic storms of cycle 24 occurred on 22-23 June 2015, i.e., during the summer solstice time. Starting from 18 June 2015, one of the biggest active regions on the Sun of current solar cycle 24 (NOAA AR 2371) produced several CMEs associated with M-class solar flares, which were observed at areas close to the central meridian. We analyzed the sequence of events, from the solar activity to FDs measurements. We investigated how multiple CMEs impacted the Earth, resulting in associated FDs. Our study was primarily based on experimental data acquired with  ground based cosmic ray detectors located in the SAA region. The main results of our analysis were as follows:

(a)
As expected, the FDs events observed by ground-based detectors were associated with the CME impacts. We found good agreement between our instruments that detected FDs. The first and most intense FD occurred on 22 June. The FD intensity (depression) with an amplitude up to 4\% was 
observed by the New-Tupi vertical telescope and Muonca (in the scaler mode) on 22 June. The El Leoncito cosmic ray detector measured an amplitude of about 3.7\%.

(b)
We found that the storm showed multi-step development reflected in several impacts (shocks) on Earth's magnetosphere. The multiple set of shocks can lead to the growth of an intense geomagnetic storm. We analyzed possible association between the schocks and the particular CME.
The first weak impact (on 21 June) was most likely caused by the 18 June full halo CME 0078. The second impact of 22 June is consistent with the 21 June CME 0090 origin. Only after the 22 June impact of the narrow CME 0091 ($pa \sim 270^{\circ}$) was G4 geomagnetic storm triggered. The 25 June G2 geomagnetic storm was triggered after the impact of CME 0093 on 23 June (fourth impact).

(c)
On the basis of the 25-26 June mini-FD event analysis we found that the forth impact (originated by CME 0093) has stretched the total duration
(from the onset time until the recovery time) of the main summer solstice FD for at least 48 h. There was about 16 h delay between the impact of the CME and the onset of the storm. This observation suggests that the fourth impact occurred under very special conditions, due to change in the Earth's space environment after the preceding CMEs.

(d) 
In this study we were interested to determine whether the storm features, basic CME properties and FDs were unusual in any way. For the purpose, we performed a cross-check between the CACTus CME catalog, geomagnetic storm parameters and the associated FD events from the ascending phase (2010-2013) of solar cycle 24.

Analysis of the pa distribution of all the CMEs that triggered FD events with an amplitude equal or above 1\% shows that there are two peaks in the equatorial region (corresponding to $pa$ 90$^{\circ}$ and 270$^{\circ}$). This result suggests that the associated CMEs mainly originate from low latitudes.

Analysis of the CMEs that triggered intense geomagnetic storms with a $Dst$ below $-70$ nT shows that there are two groups of geoeffective CMEs. The first group is well known from  previous studies, such as in \citep{gop09}. These CMEs are characterized by (1) high speed shocks, (2) a solar source location close to the central meridian and (3) a large angular width (mostly halos and partial halos). We can add that the pa distribution is wide. In this paper we showed the second group. The events show narrow width ($da <90^{\circ}$) and primarily clustered at the pa around 270$^{\circ}$. This is the case of the 21 June CME 0091.
 
Our analysis of the FDs associated with intense storms ($Dst < -70$ nT) suggests that west hemisphere CMEs ($pa$ close to 270$^{\circ}$) are more geoeffective than east hemisphere CMEs.

Using the 2010-2013 FD set, we found an event with similar characteristics ($pa\sim 289^{\circ}$ and $da\sim 76^{\circ}$). The very intense FD ($Dst\sim -133$ nT ) occurred on 8 March 2012 and was also caused by a series of CMEs. 

(e)
In this study we have investigated the 21-24 June radiation storm that reached S3 level on 22 June. We looked for a possible GLE signal in the NMDB and other ground-based detectors, including New-Tupi and Muonca detectors, but did not find the signal. We look for GLEs (or some particle excesses) associated with these CMEs in the NM network and other detectors, in all cases the result was negative.

(f)
The Sun travels along the ecliptic inclined by 23.5 degrees from the celestial equator. On the day of the Summer Solstice, the Sun is at furthest point above the celestial equator (23.5 degrees to the equinox). The largest geomagnetic storm of solar cycle 24 so far occurred near 2015 equinox (so-called St. Patrick's Day storm). The Summer solstice event gives us an opportunity to look again at the plasma and magnetic
field characteristics of CMEs that results in particular intense events evolving in the varied conditions of the Earth's space environment  

The ground level detectors within the SAA region (New-Tupi and Muonca in Brazil) are running. They have covered the maximum and the current descending phase of the cycle 24, and the analysis of other transient solar events is in progress.

\acknowledgments

This work is supported by the Conselho Nacional de Desenvolvimento Cient\'{i}fico e Tecnol\'{o}lgico (CNPq, grants \url{306605/2009-0, 312066/2016-3, 152050/2016-7, 406331/2015-4, 308494/2015-6}), Funda\c{c}\~{a}o de Amparo a Pesquisa do Estado do Rio de Janeiro (FAPERJ, grants \url{08458.009577/2011-81, E-26/101.649/2011}) and also for Funda\c{c}\~{a}o de Amparo a Pesquisa do Estado de S\~{a}o Paulo (FAPESP, grants \url{2011/50193-4, 2011/24117-9}). We are grateful to the El Leoncito CARPET detector team for observational data. We express our gratitude to the EMBRACE/INPE team, the ACE/MAG instrument team, the ACE Science Center, the CACTus catalog, the NASA GOES team and the NOAA Space Weather Prediction Center (\url{www.swpc.noaa.gov}), and the NMDB (\url{www.nmdb.eu}) for valuable information and for real time data. Special thanks are due to the  University Oulu station. Finally, we would like to thanks the referee for very valuable comments and  suggestions.


\newpage

\appendix

\begin{table}[]
\centering
\hspace{-0.0cm}
\caption{
CME parameters on the basis of CACTus CME catalog (dd.mm.yyyy,hh:mm (UT)). The geomagnetic parameter $Dst$ shows the minimum  detected value. The FD events for the period of 2010-2013 are based  on our estimations of the Oulu NM data. For these events, the sudden storm commencement (SSC) time 
is taken as the FD onset time (dd.mm.yyyy,hh:mm (UT)). The last two lines are values measured in New-Tupi telescope}
\label{my-label}
\begin{tabular}{|c|c|c|c|c|c|c|c|c|}
\hline
CME number & Date of CME occur.& pa (degree) & da (degree) & v (km s$^{-1}$) & $v_{max}$ (km s$^{-1}$) & Date FD onset & Dst (nT) & FD\% (Oulu-NM) \\ \hline
3                                                                     & 03/04/2010, 09:54                                                                      & 240                                                   & 66                                                    & 517                                                                       & 1134                                                                               & 05/04/2010, 08:26                                                                 & -81                                                & 2.5                                                      \\ \hline
31                                                                    & 26/10/2010, 00:48                                                                      & 198                                                   & 90                                                    & 140                                                                       & 267                                                                                & 30/10/2010, 10:13                                                                 & -7                                                 & 1.3                                                      \\ \hline
24                                                                    & 12/12/2010, 03:24                                                                      & 250                                                   & 82                                                    & 612                                                                       & 694                                                                                & 12/12/2010, 14:00                                                                 & -13                                                & 0.6                                                      \\ \hline
32                                                                    & 15/02/2011, 02:24                                                                      & 144                                                   & 360                                                   & 469                                                                       & 811                                                                                & 18/02/2011, 01:36                                                                 & -30                                                & 2.3                                                      \\ \hline
13                                                                    & 07/03/2011, 20:12                                                                      & 275                                                   & 220                                                   & 694                                                                       & 1644                                                                               & 10/03/2011, 06:45                                                                 & -83                                                & 2.0                                                      \\ \hline
9                                                                     & 02/06/2011, 07:24                                                                      & 99                                                    & 152                                                   & 573                                                                       & 844                                                                                & 04/06/2011, 20:45                                                                 & -39                                                & 2.3                                                      \\ \hline
40                                                                    & 07/06/2011, 06:49                                                                      & 269                                                   & 216                                                   & 694                                                                       & 1415                                                                               & 09/06/2011, 18:00                                                                 & -31                                                & 1.3                                                      \\ \hline
73                                                                    & 14/06/2011, 07:12                                                                      & 77                                                    & 184                                                   & 456                                                                       & 905                                                                                & 17/06/2011, 02:00                                                                 & -8                                                 & 1.7                                                      \\ \hline
81                                                                    & 21/06/2011, 03:16                                                                      & 84                                                    & 238                                                   & 552                                                                       & 684                                                                                & 22/06/2011, 03:00                                                                 & -25                                                & 2.5                                                      \\ \hline
35                                                                    & 09/07/2011, 00:48                                                                      & 96                                                    & 170                                                   & 473                                                                       & 657                                                                                & 11/07/2011, 09:00                                                                 & -24                                                & 1.7                                                      \\ \hline
10                                                                    & 04/08/2011, 04:12                                                                      & 358                                                   & 360                                                   & 868                                                                       & 1952                                                                               & 05/08/2011, 18:00                                                                 & -107                                               & 2.7                                                      \\ \hline
16                                                                    & 06/09/2011, 01:36                                                                      & 324                                                   & 126                                                   & 362                                                                       & 679                                                                                & 09/09/2011, 12:43                                                                 & -64                                                & 2.3                                                      \\ \hline
67                                                                    & 13/09/2011, 23:36                                                                      & 263                                                   & 32                                                    & 322                                                                       & 500                                                                                & 17/09/2011, 04:00                                                                 & -70                                                & 2.0                                                      \\ \hline
129                                                                   & 24/09/2011, 13:25                                                                      & 144                                                   & 360                                                   & 941                                                                       & 1971                                                                               & 26/09/2011, 12:37                                                                 & -101                                               & 3.0                                                      \\ \hline
4                                                                     & 01/10/2011, 09:36                                                                      & 302                                                   & 98                                                    & 351                                                                       & 500                                                                                & 05/10/2011, 08:00                                                                 & -42                                                & 1.7                                                      \\ \hline
93                                                                    & 21/10/2011, 23:36                                                                      & 309                                                   & 324                                                   & 431                                                                       & 694                                                                                & 24/10/2011, 18:00                                                                 & -132                                               & 3.5                                                      \\ \hline
119                                                                   & 27/10/2011, 11:48                                                                      & 53                                                    & 96                                                    & 522                                                                       & 722                                                                                & 01/11/2011, 08:00                                                                 & -71                                                & 1.7                                                      \\ \hline
64                                                                    & 19/01/2012, 14:36                                                                      & 310                                                   & 346                                                   & 637                                                                       & 1453                                                                               & 22/01/2012, 06:14                                                                 & -73                                                & 1.5                                                      \\ \hline
80                                                                    & 23/01/2012, 04:36                                                                      & 144                                                   & 360                                                   & 1092                                                                      & 2016                                                                               & 24/01/2012, 15:04                                                                 & -80                                                & 3.0                                                      \\ \hline
125                                                                   & 27/01/2012, 19:00                                                                      & 144                                                   & 360                                                   & 1130                                                                      & 1930                                                                               & 30/01/2012, 16:00                                                                 & -17                                                & 0.7                                                      \\ \hline
8                                                                     & 04/03/2012, 11:00                                                                      & 38                                                    & 186                                                   & 735                                                                       & 1487                                                                               & 07/03/2012, 04:21                                                                 & -78                                                & 2.0                                                      \\ \hline
18                                                                    & 07/03/2012, 01:25                                                                      & 289                                                   & 76                                                    & 947                                                                       & 1562                                                                               & 08/03/2012, 11:05                                                                 & -143                                               & 7.0                                                      \\ \hline
52                                                                    & 10/03/2012, 18:00                                                                      & 260                                                   & 218                                                   & 843                                                                       & 1837                                                                               & 12/03/2012, 09:21                                                                 & -51                                                & 3.7                                                      \\ \hline
99                                                                    & 27/05/2012, 05:48                                                                      & 85                                                    & 136                                                   & 502                                                                       & 892                                                                                & 30/05/2012, 17:00                                                                 & -5                                                 & 2.0                                                      \\ \hline
97                                                                    & 14/06/2012, 14:12                                                                      & 230                                                   & 70                                                    & 919                                                                       & 1041                                                                               & 16/06/2012, 20:00                                                                 & -86                                                & 3.0                                                      \\ \hline
6                                                                     & 01/07/2012, 15:24                                                                      & 178                                                   & 38                                                    & 601                                                                       & 1077                                                                               & 05/07/2012, 06:00                                                                 & 3                                                  & 1.5                                                      \\ \hline
34                                                                    & 04/07/2012, 17:12                                                                      & 20                                                    & 34                                                    & 351                                                                       & 512                                                                                & 08/07/2012, 04:00                                                                 & -69                                                & 2.0                                                      \\ \hline
84                                                                    & 12/07/2012, 14:24                                                                      & 86                                                    & 10                                                    & 449                                                                       & 589                                                                                & 14/07/2012, 18:11                                                                 & -133                                               & 3.5                                                      \\ \hline
120                                                                   & 19/07/2012, 06:00                                                                      & 57                                                    & 344                                                   & 1420                                                                      & 1953                                                                               & 21/07/2012, 16:00                                                                 & -21                                                & 0.7                                                      \\ \hline
150                                                                   & 31/08/2012, 20:00                                                                      & 144                                                   & 360                                                   & 644                                                                       & 1488                                                                               & 03/09/2012, 12:14                                                                 & -78                                                & 1.3                                                      \\ \hline
11                                                                    & 02/09/2012, 04:01                                                                      & 284                                                   & 158                                                   & 323                                                                       & 496                                                                                & 04/09/2012, 22:00                                                                 & -68                                                & 4.0                                                      \\ \hline
26                                                                    & 05/10/2012, 03:24                                                                      & 239                                                   & 184                                                   & 449                                                                       & 525                                                                                & 08/10/2012, 05:15                                                                 & -111                                               & 2.0                                                      \\ \hline
41                                                                    & 07/10/2012, 07:12                                                                      & 192                                                   & 194                                                   & 484                                                                       & 718                                                                                & 11/10/2012, 13:00                                                                 & -91                                                & 2.0                                                      \\ \hline
107                                                                   & 27/10/2012, 15:24                                                                      & 264                                                   & 18                                                    & 294                                                                       & 408                                                                                & 31/10/2012, 15:39                                                                 & -74                                                & 1.2                                                      \\ \hline
43                                                                    & 09/11/2012, 15:12                                                                      & 195                                                   & 200                                                   & 411                                                                       & 508                                                                                & 12/11/2012, 23:16                                                                 & -109                                               & 2.3                                                      \\ \hline
95                                                                    & 20/11/2012, 11:48                                                                      & 309                                                   & 192                                                   & 479                                                                       & 971                                                                                & 23/11/2012, 20:00                                                                 & -42                                                & 1.8                                                      \\ \hline
34                                                                    & 11/04/2013, 07:36                                                                      & 106                                                   & 278                                                   & 578                                                                       & 905                                                                                & 13/04/2013, 05:59                                                                 & -7                                                 & 3.5                                                      \\ \hline
125                                                                   & 28/04/2013, 20:48                                                                      & 236                                                   & 70                                                    & 404                                                                       & 534                                                                                & 30/04/2013, 22:54                                                                 & -67                                                & 1.0                                                      \\ \hline
10                                                                    & 03/05/2013, 18:00                                                                      & 89                                                    & 104                                                   & 568                                                                       & 771                                                                                & 05/05/2013, 16:00                                                                 & -25                                                & 2.0                                                      \\ \hline
65                                                                    & 14/05/2013, 01:25                                                                      & 18                                                    & 346                                                   & 801                                                                       & 1840                                                                               & 15/05/2013, 07:00                                                                 & -32                                                & 2.0                                                      \\ \hline
131                                                                   & 22/05/2013, 14:00                                                                      & 144                                                   & 360                                                   & 1008                                                                      & 1956                                                                               & 25/05/2013, 09:48                                                                 & -51                                                & 1.5                                                      \\ \hline
99                                                                    & 22/08/2013, 08:36                                                                      & 210                                                   & 98                                                    & 331                                                                       & 511                                                                                & 24/08/2013, 12:00                                                                 & -23                                                & 2.0                                                      \\ \hline
131                                                                   & 29/09/2013, 22:24                                                                      & 144                                                   & 360                                                   & 548                                                                       & 1953                                                                               & 02/10/2013, 02:00                                                                 & -75                                                & 1.7                                                      \\ \hline
48                                                                    & 11/10/2013, 07:36                                                                      & 80                                                    & 122                                                   & 812                                                                       & 1736                                                                               & 14/10/2013, 07:00                                                                 & -49                                                & 1.3                                                      \\ \hline
66                                                                    & 10/11/2013, 05:48                                                                      & 191                                                   & 52                                                    & 466                                                                       & 558                                                                                & 11/11/2013, 03:00                                                                 & -73                                                & 1.2                                                      \\ \hline
137                                                                   & 21/11/2013, 01:25                                                                      & 256                                                   & 74                                                    & 694                                                                       & 1162                                                                               & 22/11/2013, 22:00                                                                 & -27                                                & 1.7                                                      \\ \hline
174                                                                   & 27/11/2013, 21:28                                                                      & 109                                                   & 64                                                    & 520                                                                       & 672                                                                                & 30/11/2013, 11:00                                                                 & -28                                                & 2.0                                                      \\ \hline
50                                                                    & 12/12/2013, 03:48                                                                      & 226                                                   & 134                                                   & 694                                                                       & 1059                                                                               & 14/12/2013, 14:00                                                                 & -41                                                & 2.0                                                      \\ \hline 
91                                                                    & 21/06/2015, 02:48                                                                      & 285                                                   & 56                                                    & 1249                                                                      & 1953                                                                               & 22/06/2015, 18:37                                                                 & -200                                               & 4.0                                                    \\ \hline
93                                                                    & 22/06/2015, 18:36                                                                      & 236                                                   & 290                                                   & 702                                                                       & 1953                                                                               & 25/06/2015, 06:00                                                                 & -75                                                & 2.5                                                     \\ \hline
\end{tabular}
\end{table}

\end{document}